\documentclass{aa}
\usepackage{graphicx}

\begin{document}

%\thesaurus{3 (11.19.2;  % Galaxies: spiral
%              11.19.6;  % Galaxies: structure
%              11.16.1;  % Galaxies: photometry
%              11.09.2;  % Galaxies: interactions
%              11.06.2;  % Galaxies: fundamental parameters
%              11.11.1;  % Galaxies: kinematics and dynamics
%              11.07.1)  % Galaxies: general
%              }

\title{$BVRI$ Surface Photometry of (S+S) Binary Galaxies.}

\subtitle{I. The data\thanks{Based on data obtained at the 2.1m telescope 
of the Observatorio Guillermo Haro at Cananea, Sonora, M\'exico, operated 
by the Instituto Nacional de Astrof\'\i sica, Optica y Electr\'onica.}}

\author{H. M. Hern\'andez-Toledo\inst{1}
\and I. Puerari\inst{2}}

\offprints{H. Hern\'andez-Toledo}

\institute{Instituto de Astronom\'\i a -- UNAM -- Apartado Postal 70-264, 
04510 M\'exico D.F., M\'exico\\
\email{hector@astroscu.unam.mx}
\and Instituto Nacional de Astrof\'\i sica, Optica y Electr\'onica,
Calle Luis Enrique Erro 1, 72840 Tonantzintla, Puebla, M\'exico\\
\email{puerari@inaoep.mx}}

\titlerunning{Optical Surface Photometry of Binary Galaxies}
\authorrunning{Hern\'andez-Toledo and Puerari}
 
\date{Received .............../ Accepted August 21, 2001}

\abstract{We present multicolour broad band ($BVRI$) photometry 
for a sample of 33 spiral-spiral (S+S) binary galaxies drawn 
from the Karachentsev Catalogue of Isolated Pairs of Galaxies 
(KPG). The data is part of a joint observational programme
devoted to systematic photometric study of one of the most 
complete and homogeneous pair samples available in the 
literature. We present azimuthally averaged colour and surface 
brightness profiles, colour index ($B-I$) maps, $B$ band and 
sharp/filtered $B$ band images as well as integrated 
magnitudes, magnitudes at different circular apertures 
and integrated colours for each pair. Internal and external
data comparisons show consistency within the estimated 
errors. Two thirds of the sample have total aperture 
parameters homogeneously derived for the first time. After 
reevaluating morphology for all the pairs, we find a change
in Hubble type for 24 galaxies compared to the original
POSS classifications. More than half of our pairs show 
morphological concordance which could explain, in part, 
the strong correlation in the ($B-V$) colour indices 
(Holmberg Effect) between pair components. We find
a tendency for barred galaxies to show grand design 
morphologies and flat colour profiles. The measurements 
will be used in a series of forthcoming papers where we try
to identify and isolate the main structural and photometric 
properties of disk galaxies at different stages of interaction. 
}

\maketitle

\section{Introduction}

Two-dimensional broad-band photometry has been systematically 
applied to the study of binary galaxy structure and dynamics, 
only in the past two decades. The mid to late 70's saw an 
astronomical debate that led to the recognition that 
gravitational interaction is an important factor in galactic 
evolution affecting directly properties such as size, 
morphological type, luminosity, star formation rate, and 
mass distribution (Sulentic \cite{sulentic76}; Larson \& Tinsley 
\cite{larsontinsley78}; Stocke \cite{stocke78}). According to 
current popular models of galaxy formation, galaxies are assembled 
through a hierarchical process of mass aggregation, dominated 
either by mergers (Kauffmann, White \& Guiderdoni 
\cite{kauffmannetal93}; Baugh, Cole \& Frenk \cite{baughetal96}) 
or by gas accretion (Avila-Reese, Firmani \& Hern\'andez 
\cite{avilaetal98}; Avila-Reese \& Firmani \cite{avilafirmani00}). 
In the light of these models, the influence of environment factors 
and interaction phenomena in the shaping and star formation of the 
disks is natural, at least for a fraction of the present-day 
galaxy population. Examples are galaxy harassment in clusters 
(Moore et al. \cite{mooreetal96}), tidal stirring of dwarf 
irregulars near giant galaxies (Mayer et al. \cite{mayeretal01}), 
tidally induced star formation (Lacey \& Silk 
\cite{laceysilk91}; Kauffmann, Charlot \& Balogh
 \cite{kauffmannetal01}). See also Dultzin-Hacyan (1997) for a 
non-biased review.

For binary galaxies, current ideas suggest that most 
physical pairs are morphologically concordant, that is, 
with components showing similar initial star formation and 
angular momentum properties. Evidence in favour of these ideas
come from the colour correlations (Holmberg effect) observed 
for components of pairs, although very few photometric data 
supporting this correlation exists in the present. The large 
number of (S+S) pairs in The Catalogue of Isolated Pairs of 
Galaxies in the  Northern Hemisphere (KPG, Karachentsev 
\cite{kara72}) means that for a flux limited sample 
($m_{zw} = 15.7$), almost six out of every ten pairs are 
of the (S+S) type, suggesting that a considerable number 
of them must be physical binaries.

Redshift information, available for the whole (S+S) 
sample, suggests that most of them are likely to be physically 
proximate. Digital Sky Survey images show that most of them 
have visible signs of disturbance; bridges, tails, common
envelopes and distortions that are regarded as evidence
for gravitational interaction. In addition, statistical 
studies indicate that a high fraction ($\sim 65$\%) 
show an enhancement in the optical and FIR emission 
(Xu \& Sulentic \cite{xusulentic91}; Hern\'andez-Toledo et 
al. \cite{toledoetal99}). This enhancement is interpreted as 
a by-product of interaction-induced star formation activity 
in physical binaries.

One of the most important lessons that emerges from 
statistical studies of interacting galaxies is that if we 
want to have a better understanding of the connection between 
interactions, photometric structural parameters 
and optical morphology, it is necessary to obtain accurate
photometry for complete and homogeneous interacting galaxy 
samples. The results can then be interpreted by applying similar 
methods to well-matched comparison samples (isolated, presumably 
undisturbed galaxies). Some efforts in this direction 
are Bergvall \& Johansson (\cite{bergvalljohansson95}), 
Reduzzi \& Rampazzo (\cite{reduzzirampazzo96}), M\'arquez 
\& Moles (\cite{marquezmoles96}), Laurikainen et al. 
(\cite{laurikainenetal98}), Jansen et al. (\cite{jansenetal00}),
M\'arquez \& Moles (\cite{marquezmoles99}), and de Jong \& van der
Kruit (\cite{dejongvanderkruit94}).

Our main goal is to obtain an homogeneous set of 
broad-band observations for most (if not all) of the (S+S) 
pairs in the Karachentsev Catalogue. The relative simplicity, 
compared to groups and clusters along with the size, brightness 
limit and morphological diversity, offer us a unique opportunity 
to realize accurate photometric observations for a statistically 
significant number of pairs where a less confused interpretation 
of the photometric properties of disk galaxies at different 
stages of interactions (and its relationship to optical morphology)
is possible.

We present in this first paper of a series, our photometric
data with emphasis on the morphological properties in a subset 
of 33 (S+S) pairs. The structure of the paper is as follows. 
Section \ref{sample} summarizes some limitations and biases 
in the (S+S) sample which are relevant to our photometric 
study. Section \ref{datareduction} presents the observations,
reduction techniques and a comparison of our estimated
total magnitudes against those in the literature. A discussion 
on the related errors is also included. Section \ref{discussion} 
shows a discussion based on the estimated colour indices, 
observed optical morphologies and (colour and surface brightness) 
profiles for each component galaxy. The systematics of 
morphological distortions induced by the interactions are 
commented in the light of current models. Section 
\ref{conclusion} is a summary of the conclusions achieved.
Finally, an appendix is devoted to the
presentation of magnitudes at three different 
concentric circular apertures.

\section{Sample and Observations}
\label{sample}

Through the main optical observatories in M\'exico (Observatorio
Astron\'omico Nacional at San Pedro M\'artir in Baja California 
and Observatorio Astrof\'\i sico Guillermo Haro in Cananea) 
we have started a joint observational programme devoted 
to obtain uniform photometric data for one of the most complete 
and homogeneous pair samples currently available. The sample of 
disk-disk (S+S) pairs amounts to more than 300 pairs from 
a total of 602 pairs in the KPG catalogue. The observations
were begun since January 1999. The CCD $BVRI$ images reported 
here ({\bf in the Cousins system}) were obtained with a LFOSC 
detector attached to the 2.1m telescope, at Observatorio 
Guillermo Haro, Cananea, Sonora, M\'exico, covering an area 
of about $6\arcmin \times 4\arcmin$, with a scale of 
1$\arcsec$/pixel.

Since our goal is to observe all or most of the KPG (S+S)
sample, we have applied no special strategy in selecting the 
current subset of 33 (S+S) pairs. Available observing time and 
weather conditions, were the main factors limiting the number 
of observed pairs. This will be the observational strategy 
for the next reports up to the point where most of the (S+S)
sample is observed. The selection criteria and statistical
properties for the (S+S) sample that are most relevant 
to the present and further photometric analysis are stated 
here.

\subsection{Statistical Properties of the (S+S) Sample}
 
The isolated (S+S) pairs in the KPG sample were selected from 
a visual search of the Palomar Sky Survey. The catalogue samples
the sky north of $\delta \geq -3\degr$. The vast majority of 
objects are found in high Galactic latitude regions ($b \geq 
20\degr$) and as a sample, they are reasonably complete 
($\sim 90$\%) in the magnitude range $13.5 \leq m_{zw} \leq 
15.7$. All galaxies in the (S+S) sample have measured redshifts. 
We next summarize possible limitations and sources of bias 
(mainly due to the optical selection criteria) that may affect 
the interpretation of the photometric analysis.

\begin{description}
\item{(1)} Although the projected physical separation 
(H$_0= 75$ km s$^{-1}$ Mpc$^{-1}$) for the whole (S+S) 
sample is small ($<x>$ = 42.1 kpc), there is still a source 
of contamination from accordant redshift optical pairs. 
This is a difficult source to evaluate because most such 
optical pairs are members of loose group structures with 
magnitude and redshift properties completely within the 
domain of expectation for physical binary systems. 
Our photometric study, however, is the most helpful to
try to solve the problem if we observe as much as 
possible the fraction of the (S+S) sample that shows 
direct/indirect signs of interactions. That is, by observing 
the most probable physical pairs, on an individual basis.

\item{(2)} The (S+S) sample, will reject highly evolved pairs 
such as mergers in the final stages of coalescence. This 
stems from the selection requirement that the galaxies have 
a discernible diameter. This excludes, for example, 
ultra-luminous infrared galaxies (ULIRGs)
from our photometric study of (S+S) pairs.

\item{(3)} The maximum size ratio between components (from the 
basic selection criteria) is $ a_{2}/a_{1} \sim 8$.
This means that the (S+S) sample favours magnitude and thus size
concordance, biasing this sample against hierarchical
binaries. However, an important difference in visual apparent
magnitudes between member pairs ($\Delta m \sim 3^{m}$) can be
found.

\item{(4)} (S+S) pairs with the faintest apparent magnitudes
must have smaller apparent separations in order to survive the 
isolation criterion and enter the Karachentsev catalogue.
This means that apparently close pairs are biased towards 
physically close binaries -- pairs near pericenter  (van Moorsel
 \cite{vanmoorsel82}), or with smaller mean physical separations. 
By selecting the brightest (S+S) pairs, we therefore sample a 
wider range of physical separations in our photometric study.

\item{(5)} The (S+S) pairs were selected whith a strong
isolation criteria. Thus, we expect that only intrinsic 
properties of the individual galaxies and the effects of their 
mutual interactions should affect the observed morphological and 
photometrical properties.
\end{description}

\section{Observations and Data Reduction}
\label{datareduction}

A journal of the first set of the photometric observations is 
given in Table \ref{tabjournal}. Column (1) gives the original 
catalogued number, Columns (2)-(9) give the number of frames per 
filter, the integration time (in seconds) and seeing conditions 
(in arcsec).

\begin{table*}
  \caption[]{Journal of observations. The number of frames per
filter, the integration time (in seconds), and the mean FWHM for 
each observation (in arcsec) are given.}

\begin{tabular}{ccccccccc}

Galaxy pair & $B$ & $<B>_{\rm FWHM}$ & $V$ & $<V>_{\rm FWHM}$ & $R$ & $<R>_{\rm 
FWHM}$ & $I$ & $<I>_{\rm FWHM}$\\
KPG64  & 2$\times$1200 &2.4& 3$\times$180  &2.3& 5$\times$90   &2.6& 5$\times$90
&3.1  \\
KPG68  & 1$\times$1200 &2.3& 5$\times$180  &2.5& 10$\times$60  &2.3& 5$\times$120
&2.8  \\
KPG75  & 1$\times$1800 &2.1& 2$\times$600  &2.2& 3$\times$240  &2.1& 3$\times$180
&2.7  \\
KPG88  & 1$\times$1800 &2.4& 1$\times$1800 &2.5& 1$\times$1200 &2.6& 1$\times$900
&3.3  \\
KPG98  & 1$\times$1800 &2.6& 1$\times$900  &2.5& 2$\times$300  &3.1& 2$\times$300
&3.1  \\
KPG102 & 1$\times$1800 &2.1& 1$\times$1200 &2.4& 1$\times$600  &2.4& 1$\times$600
&2.6  \\
KPG103 & 1$\times$1800 &2.7& 1$\times$600  &2.5& 1$\times$300  &2.5& 1$\times$300
&2.8  \\
KPG108 & 1$\times$1800 &2.5& 1$\times$1200 &2.5& 1$\times$600  &2.2& 1$\times$300
&2.4  \\
KPG112 & 1$\times$1800 &2.5& 3$\times$180  &2.3& 5$\times$60   &2.2& 5$\times$60
&2.8  \\
KPG125 & 1$\times$1800 &2.6& 1$\times$1200 &2.9& 2$\times$300  &3.2& 2$\times$300
&3.6  \\
KPG136 & 1$\times$1800 &2.7& 2$\times$900  &2.5& 1$\times$600  &2.3& 2$\times$300
&2.9  \\
KPG141 & 1$\times$1800 &2.4& 1$\times$1200 &2.3& 1$\times$600  &2.2& 1$\times$600
&2.5  \\
KPG150 & 1$\times$1800 &2.7& 2$\times$600  &2.4& 2$\times$300  &2.4& 3$\times$180
&2.7  \\
KPG151 & 1$\times$1800 &2.4& 1$\times$1200 &2.6& 1$\times$900  &2.2& 1$\times$600
&3.3  \\
KPG156 & 1$\times$1800 &2.6& 1$\times$1200 &2.6& 1$\times$600  &2.9& 1$\times$600
&3.6  \\
KPG159 & 1$\times$1800 &3.6& 1$\times$600  &3.4& 1$\times$600  &3.2& 1$\times$300
&3.5  \\
KPG160 & 1$\times$1800 &2.5& 2$\times$600  &2.1& 3$\times$240  &2.3& 3$\times$240
&2.8  \\
KPG168 & 1$\times$1200 &3.1& 2$\times$300  &2.9& 3$\times$120  &2.9& 3$\times$120
&3.5  \\
KPG195 & 1$\times$1200 &3.0& 2$\times$300  &3.2& 5$\times$90   &2.6& 5$\times$90
&2.5  \\
KPG211 & 1$\times$1800 &2.7& 1$\times$600  &2.5& 2$\times$240  &3.0& 3$\times$180
&3.7  \\
KPG216 & 1$\times$1800 &2.9& 1$\times$1200 &2.7& 1$\times$600  &2.3& 1$\times$600
&2.4  \\
KPG249 & 1$\times$1800 &3.0& 1$\times$600  &3.3& 1$\times$300  &3.5& 1$\times$300
&3.4  \\
KPG295 & 1$\times$1800 &3.0& 2$\times$600  &3.2& 5$\times$150  &3.1& 5$\times$120
&2.9  \\
KPG302 & 1$\times$1200 &2.3& 3$\times$300  &2.3& 5$\times$180  &2.2& 6$\times$120
&2.7  \\
KPG313 & 1$\times$1800 &3.3& 1$\times$600  &3.2& 3$\times$180  &3.1& 3$\times$120
&2.6  \\
KPG332 & 1$\times$1800 &2.9& 1$\times$1200 &2.9& 2$\times$300  &2.5& 2$\times$240
&2.6  \\
KPG347 & 1$\times$900  &2.8& 1$\times$600  &3.0& 3$\times$180  &2.5& 3$\times$120
&2.6  \\
KPG389 & 1$\times$1800 &3.5& 1$\times$1200 &3.6& 2$\times$600  &3.2& 2$\times$300
&2.6  \\
KPG396 & 1$\times$1800 &2.4& 1$\times$1200 &2.7& 1$\times$900  &2.5& 1$\times$900
&2.7  \\
KPG404 & 1$\times$900  &2.7& 2$\times$300  &2.9& 5$\times$150  &3.1& 5$\times$120
&2.4  \\
KPG426 & 1$\times$1800 &2.9& 1$\times$600  &3.1& 2$\times$300  &3.2& 3$\times$180
&3.0  \\
KPG440 & 1$\times$1200 &2.4& 1$\times$900  &3.0& 2$\times$300  &2.8& 2$\times$240
&2.6  \\
KPG455 & 1$\times$1800 &2.3& 1$\times$600  &2.5& 3$\times$180  &2.6& 3$\times$180
&3.1  \\

\label{tabjournal}
\end{tabular}
\end{table*}

Table \ref{tabgeneraldata} reports some relevant information
for the observed pairs coming from the literature. Column 
(1) is the KPG catalogued number, Column (2) reports other
identifications, Column (3) the apparent $B$ magnitude from
the Nasa Extragalactic Database (NED), Column (4) the linear
separation (in kpc), Column (5) the radial velocity in
km s$^{-1}$ from NED, and finally, Column (6) gives the major
axis diameter (at $\mu_{B} = 25$) for each component galaxy
(in kpc).

\begin{table*}
  \caption[]{General Data for the Observed Galaxies}
\label{tabgeneraldata}
\begin{tabular}{ccccccc}
\hline\noalign{\smallskip}
KPG Number & Identif. & $B$ mag & $x_{12}$ (kpc) & $V_{Rad}$ (km s$^{-1}$)
& $A_{25}$ (kpc) \\
\hline\noalign{\smallskip}
KPG64A  & UGC01810 & 13.42(p) & 39.5 & 7563 & 55.8 \\
KPG64B  & UGC01813 & 15.08(p) &      & 7335 & 27.4 \\
KPG68A  & NGC0935  & 13.63(p) & 17.4 & 4142 & 29.3 \\
KPG68B  & IC1801   & 14.56(p) &      & 4023 & 19.3 \\
KPG75A  & UGC02222 & 14.56(p) & 21.5 & 4913 & 32.0 \\
KPG75B  & UGC02225 & 15.21(p) &      & 4965 & 18.6 \\
KPG88A  & UGC02627 & 14.89(a) & 32.7 & 4224 & 30.8 \\
KPG88B  & UGC02629 & 15.28(p) &      & 4128 & 15.1 \\
KPG98A  & UGC02954 & 15.23(p) & 41.2 & 5306 & 17.4 \\
KPG98B  & MRK1081  & 15.15(p) &      & 5345 & 16.0 \\
KPG102A & CGCG393-070 & 15.50(p) & 34.4 & 10778 & 21.0 \\
KPG102B & UGC03136 & 15.00(p) &        & 10674 & 34.8 \\
KPG103A & CGCG420-003 & 15.70(p) & 55.2 & 8313  & 20.3 \\
KPG103B & UGC03179 & 14.46(p) &        & 8337  & 28.4 \\
KPG108A & UGC03405 & 15.32(p) & 33.9 & 3738 & 20.1 \\
KPG108B & UGC03410 & 14.99(p) &      & 3921 & 29.8 \\
KPG112A & UGC03445 & 14.25(p) & 9.7  & 3119 & 21.3 \\
KPG112B & UGC03446 & 13.86(p) &      & 3116 & 21.2 \\
KPG125A & NGC2341  & 13.84(a) & 50.9 & 5227 & 24.9 \\
KPG125B & NGC2342  & 13.10(a) &      & 5276 & 29.1 \\
KPG136A & CGCG086-028 & 14.80(p) & 37.8 & 9907 & 31.1 \\
KPG136B & CGCG086-029 & 15.00(p) &      & 9813 & 32.1 \\
KPG141A & UGC04005 & 14.60(p) & 89.8 & 5044 & 31.8 \\
KPG141B & CGCG030-014 & 14.80(p) &   & 4896 & 14.4 \\
KPG150A & NGC2486  & 14.16(a) & 99.3 & 4649 & 29.5 \\
KPG150B & NGC2487  & 13.23(a) &      & 4841 & 45.6 \\
KPG151A & UGC04133 & 16.00(p) & 32.5 & 9130 & 50.5 \\
KPG151B & UGC04134 & 15.37(p) &      & 8968 & 32.6 \\
KPG156A & NGC2535  & 13.31(a) & 27.7 & 4097 & 30.9 \\
KPG156B & NGC2536  & 14.70(a) &      & 4142 & 17.0 \\
KPG159A & CGCG088-052 & 15.60(p) & 26.0 & 5232 & 9.9 \\
KPG159B & UGC04286 & 14.32(p) &         & 5143 & 19.9 \\
KPG160A & NGC2544  & 13.80(a) & 17.4 & 2828 & 16.0 \\
KPG160B & CGCG331-037 & 15.50(p) &  & 3589 & 11.0 \\
KPG168A & NGC2648  & 12.74(p) & 17.6 & 2060 & 23.1 \\
KPG168B & CGCG060-036 & 15.40(p) &   & 2115 & 9.2 \\
KPG195A & NGC2798  & 13.04(a) & 11.5 & 1726 & 16.3  \\
KPG195B & NGC2799  & 14.32(p) &      & 1865 & 11.3  \\
KPG211A & NGC2959  & 13.65(p) & 26.8 & 4429 & 29.3  \\
KPG211B & NGC2961  & 15.52(p) &      & 4501 & 12.2  \\
KPG216A & NGC3018  & 14.13(p) & 17.8 & 1863 & 6.45 \\
KPG216B & NGC3023  & 13.50(p) &      & 1879 & 14.7 \\
KPG249A & NGC3395  & 12.40(a) & 8.8  & 1625 & 10.5 \\
KPG249B & NGC3396  & 12.63(p) &      & 1625 & 12.3  \\
KPG295A & NGC3786  & *13.24(p) & 14.9 & 2678 & 19.2 \\
KPG295B & NGC3788  & 13.46(p) &      & 2699 & 13.9  \\
KPG302A & NGC3893  & 11.16(s) & 13.9 & 977  & 17.1 \\
KPG302B & NGC3896  & 13.89(p) &      & 980  & 5.9  \\
KPG313A & IC0749   & 12.92(s) & 10.9 & 784  & 7.5 \\
KPG313B & IC0750   & 12.94(s) &      & 701  & 9.1 \\
KPG332A & NGC4298  & 12.04(s) & 9.5  & 1135 & 10.3 \\
KPG332B & NGC4302  & 12.50(s) &      & 1149 & 13.0 \\
KPG347A & NGC4567  & 12.06(s) & 10.8 & 2274 & 20.2 \\
KPG347B & NGC4568  & 11.68(s) &      & 2255 & 29.3  \\
\hline\noalign{\smallskip}
\end{tabular}
\end{table*}

\setcounter{table}{1}

\begin{table*}
  \caption[]{Continued.}

\begin{tabular}{ccccccc}

\hline\noalign{\smallskip}
KPG Number & Identif. & $B$ mag & $x_{12}$ (kpc) & $V_{Rad}$ (km s$^{-1}$)
& $A_{25}$ (kpc) \\
\hline\noalign{\smallskip}
KPG389A & NGC5257  & *13.50(p) & 36.6 & 6798 & 39.0 \\
KPG389B & NGC5258  & *13.49(p) &      & 6757 & 37.8 \\
KPG396A & UGC08713 & 15.25(p) & 28.6 & 4956 & 24.3 \\
KPG396B & UGC08715 & 14.50(p) &      & 4517 & 22.7 \\
KPG404A & NGC5394  & 13.70(a) & 26.2 & 3472 & 22.8 \\
KPG404B & NGC5395  & 12.10(a) &      & 3491 & 34.1 \\
KPG426A & UGC09376 & 14.70(p) & 26.8 & 7676 & 40.6  \\
KPG426B & CGCG220-030 & 14.89(p) &   & 7764 & 46.8  \\
KPG440A & NGC5774  & 12.74(s) & 27.2 & 1567 & 20.9 \\
KPG440B & NGC5775  & 12.24(s) &      & 1681 & 21.2  \\
KPG455A & NGC5857  & 13.86(a) & 38.4 & 4682 & 21.6  \\
KPG455B & NGC5859  & 13.27(a) &      & 4764 & 41.7 \\
\hline\noalign{\smallskip}
\end{tabular}

(a) total (asymptotic) magnitude in the $B$ system, derived
by extrapolation from\\
photoelectric aperture-magnitude data.\\
(s) total asymptotic magnitude in the $B$ system, derived
by extrapolation from\\
(surface) photometry with photoelectric zero point.\\
(p) photographic magnitude reduced to the $B_{T}$ system.
\end{table*}

Images were debiased, trimmed, and flat-fielded using
standard IRAF\footnote{The IRAF package is written and supported 
by the IRAF programming group at the National Optical Astronomy
Observatories (NOAO) in Tucson, Arizona. NOAO is operated by the 
Association of Universities for Research in Astronomy (AURA), Inc. 
under cooperative agreement with the National Science Foundation 
(NSF).} procedures. First, the bias level of the CCD was subtracted 
from all exposures. A run of 5-10 bias images was obtained per 
night, and these were combined into a single bias frame which was 
then applied to the object frames. The images were flat-fielded 
using sky flats taken in each filter at the beginning and/or at 
the end of each night.

Photometric calibration was achieved by nightly 
observations of standard stars of known magnitudes 
from the ``Dipper Asterism'' M67 star cluster
(Chevalier \& Ilovaisky \cite{chevalierilovaisky91}).
A total of 29 standard stars with a colour range $ -0.1
\leq (B-V) \leq 1.4$ and a similar range in $(V-I)$ were 
observed. The principal extinction coefficients in $B$, $V$, 
$R$ and $I$ as well as the colour terms were calculated 
according to the following equations:

%\begin{equation}
$$B-b = \alpha_{B} + \beta_{B}(b-v)_{0}$$
$$V-v = \alpha_{V} + \beta_{V}(b-v)_{0}$$
$$R-r = \alpha_{R} + \beta_{R}(v-r)_{0}$$
$$I-i = \alpha_{I} + \beta_{I}(v-r)_{0}$$
%\end{equation}

\noindent where $B$, $V$, $R$ and $I$ are the standard 
magnitudes, $b$, $v$, $r$ and $i$ are the instrumental (and 
airmass-corrected) magnitudes. $\alpha$ and $\beta$ are the 
transformation coefficients for each filter.

In a first iteration, a constant value associated with the 
sky background was subtracted using an interactive procedure
that allows the user to select regions on the frame free
of galaxies and bright stars. However, occasionally, at the 
end of the reduction procedure, we still had images with a 
noticeable gradient in the sky background. For these images, 
a fifth-order polynomial was fitted and subtracted from the 
entire frame. After this processing , the sky background is 
usually flat to a level $\sim 1-2$\%. Errors in determining
the sky background, are, in fact, probably the dominant 
source of error in the estimation of the colour and surface 
brightness profiles. For this reason, we decided to apply 
this polynomial correction to all the images in this work.

The most energetic cosmic-ray events were automatically
masked using the COSMICRAYS task and field stars were
removed using the IMEDIT task when necessary. Within the 
galaxy itself, care was taken to identify superposed stars.
A final step in the basic reduction involved registration of 
all available frames for each galaxy and in each filter to 
within $\pm 0.1$ pixel. This step was performed by measuring 
centroids for foreground stars on the images and then performing 
geometric transformations using GEOMAP and GEOTRAN tasks in 
IRAF.

Elliptical surface brightness contours were fitted using
the STSDAS package ISOPHOTE. An initial starting guess for 
the ellipse-fitting routine was provided interactively by 
estimating points that represent the ends of the major 
and minor axis at an isophotal level of relatively high 
signal-to-noise ratio. Since we are interested on the mean 
global properties of these profiles and not in their 
detailed structure, we report azimuthally averaged profiles 
for spirals by fitting ellipses with a fixed position 
angle and ellipticity previously determined on the external 
isophotes of each galaxy. A more detailed analysis and 
interpretation will be presented in a forthcoming paper 
(Hern\'andez-Toledo \& Puerari, in preparation).

\subsection{Errors}

Total magnitudes can be calculated by analytically 
extrapolating a fitting of a disk beyond the 
outermost isophote to infinity. However, disk fitting is
notoriously fraught with uncertainty (c.f. Knapen \& van 
der Kruit \cite{knapenkruit91}). Alternatively, we 
estimate in this work a total magnitude computed from
polygonal apertures chosen interactively to assure that they
are large enough to contain the whole galaxy and still small
enough to limit the errors due to the sky error and light
contamination from a neighbor galaxy. This is achieved in
each band by using polygonal apertures with and without 
the the sky background removed within POLYPHOT routines in 
IRAF. In an appendix, we are also reporting total magnitudes 
at three different circular apertures by using the PHOT 
routines in IRAF. Foreground stars within the aperture were 
removed interactively. In some cases, the separation of the 
galaxies allowed us to model the light distribution in each 
galaxy and then to try an iterative subtraction as reported 
in Junqueira et al. (\cite{junquieraetal98}).
In cases where this procedure was not possible, our 
estimations must be taken with care. See Table 
\ref{tabmagcolours} and comments on individual objects.

An estimation of the errors in our photometry involves
two parts: 1) The procedures to obtain instrumental 
magnitudes and 2) the uncertainty when such instrumental 
magnitudes are transformed to the standard system. For 1),
notice that the magnitudes produced at the output of the IRAf
routines (QPHOT, PHOT and POLYPHOT) have a small error
that is internal for those procedures. Since we also have 
applied extinction corrections to the instrumental magnitudes
in this step, our estimation of the errors are mainly 
concerned with these corrections and the estimation of the 
airmass. After a least square fitting, the associated errors 
to the slope for each principal extinction coefficient
are; $\delta(k_{B}) \sim 0.038$, $\delta(k_{V}) \sim 0.035$, 
$\delta(k_{R}) \sim 0.020$ and $\delta(k_{I}) \sim 0.020$. 
An additional error $\delta(airmass) \sim 0.005$ from the 
airmass routines in IRAF was also considered.

For 2), the zero point and first order colour
terms are the most important to consider. After
transforming to the standard system, by adopting our 
best-fit coefficients, the formal errors from the assumed 
relations for $\alpha$  were 0.05, 0.04, 0.04 and 0.04 in
$B$, $V$, $R$ and $I$ and 0.04, 0.03, 0.03 and 0.04 for 
$\beta$. To estimate the total error in each band, it is 
necessary to use the transformation equations and then 
propagate the errors. Total typical uncertainties are 
0.15, 0.14, 0.15 and 0.14 in $B$, $V$, $R$ and $I$ bands, 
respectively.

The estimated total magnitudes in this work were compared 
against other external estimations reported in the literature.
This has been done for: 1) The standard stars and 2) those
paired galaxies in common with other works.

\subsection{Standard Stars} 

\begin{figure}
%\resizebox{\hsize}{!}{\includegraphics{m67_1_e17.eps}}
\resizebox{\hsize}{!}{\includegraphics{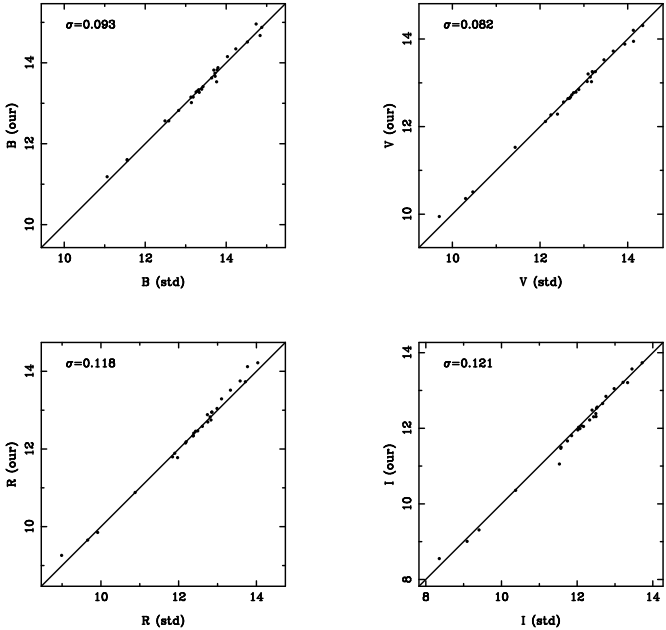}}
\caption{Comparison between our estimated magnitudes
and those from Chevalier \& Ilovaisky 1991 for 29 standard
stars in common}
\label{fig_mag_our_std}
\end{figure}

For the standard stars, a comparison of our CCD magnitudes
against those reported in Chevalier \& Ilovaisky
(\cite{chevalierilovaisky91}) for 29 stars in common, are
shown in Figure \ref{fig_mag_our_std}.

Figure \ref{fig_mag_our_std} shows no significant deviations
between our CCD magnitudes and the standard star magnitudes.
According to these results, a $\sigma \sim 0.13$, or a
similar value, could be expected as the typical error for our 
magnitude estimations in paired galaxies. This is in agreement 
with our error estimations.

\subsection{Paired Galaxies}

\begin{figure}
%\resizebox{\hsize}{!}{\includegraphics{fig_rc3_our.ps}}
\resizebox{\hsize}{!}{\includegraphics{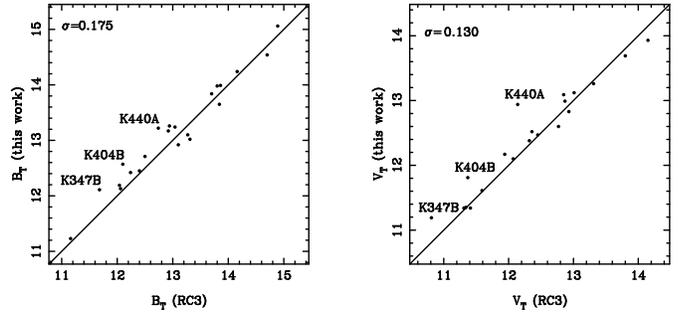}}
\caption{Comparison between our total $B$ and $V$ magnitudes 
and total magnitudes from RC3 Catalogue.}
\label{fig_mag_our_rc3}
\end{figure}

We begin with a comparison in Figure \ref{fig_mag_our_rc3}
of our total magnitudes in $B$ and $V$ bands
and those reported in the RC3 Catalogue (de Vaucouleurs 
et al. \cite{devaucouleursetal91}). 

We find that, except for three galaxies (KPG347B, 
KPG404B and KPG440A) the agreement with our measures is 
reasonably good. RMS values from our comparisons  
are 0.17 and 0.13 mag in $B$ and $V$ bands respectively.
However, as noted in Table \ref{tabmagcolours}
KPG347 and KPG404 involve two overlapping pairs (CP)
where our iterative magnitude estimation procedure could 
produce some error. In addition, the associated errors in 
$B$ and $V$ magnitudes reported in RC3 are 0.1 mag
for KPG347B and KPG440A and 0.2 mag for KPG404B.

It is important to note that a high fraction of the 
RC3 data available for our pairs comes from Zwicky 
photographic magnitudes that were transformed to the 
$B_{T}$ system ((p) in Table \ref{tabgeneraldata}). 
The possibility of systematic errors in the Zwicky 
magnitudes has been discussed by numerous authors 
(cf. Haynes \& Giovanelli \cite{haynesgiovanelli84}). 
Although these and other authors present recursion 
relations to convert Zwicky magnitudes to those of 
other systems, most notably to the photographic 
magnitudes in the Holmberg system (\cite{holmberg58}), 
or to the $B_{T}$ system (de Vaucouleurs et al. 
\cite{devaucouleursetal91}), it has been shown that 
these recursion relations are probably unsatisfactory 
for magnitudes fainter than 14.0. For this reason, in 
Figure \ref{fig_mag_our_rc3} we take into account only 
the total (asymptotic) magnitudes in RC3 derived by 
extrapolation either from photoelectric aperture-magnitude 
data ((a) in Table \ref{tabgeneraldata}) or from surface 
photometry with photoelectric zero point ((s) in Table
\ref{tabgeneraldata}). 
 
In relation to this comparison, Reshetnikov 
(\cite{reshetnikov93}) reports that the total 
magnitudes for interacting galaxies in RC3 obtained 
by means of photographic photometry are $\sim 0.2-0.3$ 
brighter compared to magnitudes from surface photometry 
with photoelectric zero point or by extrapolating the 
photoelectric data.

\begin{figure}
%\resizebox{\hsize}{!}{\includegraphics{compara_mags.ps}}
\resizebox{\hsize}{!}{\includegraphics{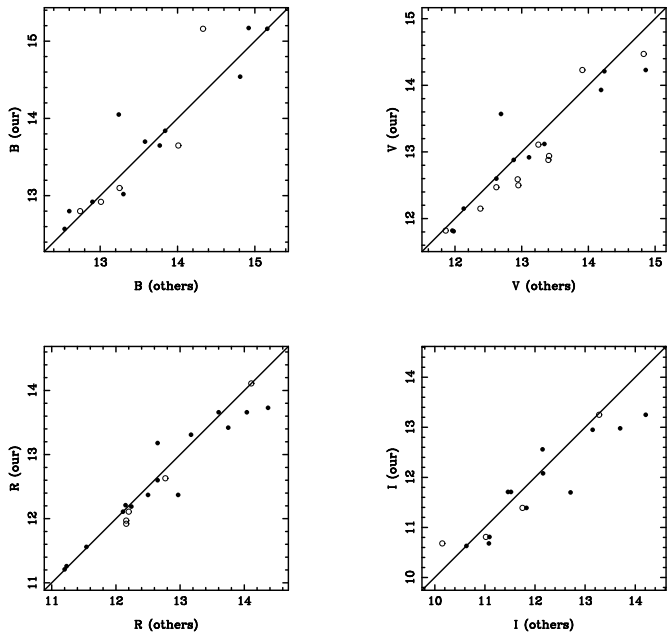}}
\caption{Comparison between our CCD data and available CCD
data from various authors for (S+S) galaxies.}
\label{compara_CCD_mags}
\end{figure}

The next step involves comparison of our CCD magnitudes 
with other CCD measures in the four colour bands. Figure
\ref{compara_CCD_mags} shows the comparison with
filled symbols denoting CCD measurements in the Cousins 
system, primarily from Han (\cite{han92}); Reshetnikov 
(\cite{reshetnikov93}); and Laurikainen et al. 
(\cite{laurikainenetal98}). Open symbols denote CCD 
measurements in the Johnson system, mainly Godwin et al. 
(\cite{godwinetal77}); Doroshenko \& Terebizh 
(\cite{doroshenkoterebizh79}); de Vaucouleurs \& Longo
(\cite{devaucouleurslongo88}) and M\'arquez \& Moles 
(\cite{marquezmoles96}). Metcalfe et al. (\cite{metcalfeetal98}) 
reports $B$ and $V$ band photometry in the Landolt system 
while $R$ and $I$ are in the Cousins system. Giovanelli et al.
(\cite{giovanellietal97}) report $I$-band data from a
combination of sources. No attempt has been made to transform 
from any of the above photometric systems to Cousins system.

Notice that in Figure \ref{compara_CCD_mags} there seems to
be no clear systematic tendency between the compared data,
in spite of the small number of galaxies in common.
The sigma values obtained through a comparison (only) in 
the Cousins systems are 0.25, 0.25, 0.20 and 0.30 in $B$, $V$, 
$R$ and $I$ respectively. However, it is fair to mention that 
this is not a straightforward comparison, since
we are also comparing both intrinsic and extrinsic
differences involved in each photometric system as well as
differences in the reduction procedures, that are more
easily detected at fainter magnitudes.

Finally, for most paired galaxies, more than one long exposure 
per filter is available. Thus we evaluate in addition, the
internal accuracy of our photometry by comparing the total 
magnitudes derived from the individual exposures. We find RMS 
differences between individual measurements of $\delta(B) \sim 
0.06$, $\delta(V) \sim 0.06$, $\delta(R) \sim 0.05$ and $\delta(I) 
\sim 0.05$. Additionally, by estimating total magnitudes for all
galaxies before and after sky subtraction, typical values 
$\delta(B) \sim 0.06$, $\delta(V) \sim 0.07$, $\delta(R) 
\sim 0.06$ and $\delta(I) \sim 0.07$ are obtained. In the
appendix, we report additional estimations of magnitudes at
three concentric circular apertures for all the paired galaxies
in this study.

\section{Discussion}
\label{discussion}

\subsection{Integrated Magnitudes and Colours}

The estimated magnitudes and colours of the galaxies in 
the sample are presented in Table \ref{tabmagcolours}.
Entries are as follows: Column (1) gives the identification 
in Karachentsev Catalogue, Columns (2) to (5) give the
observed total integrated magnitudes in $B$, $V$, $R$ and
$I$ bands, Columns (6) and (7) give the observed $(B-V)$ and
$(B-I)$ colour indices. Finally Columns (8) and (9) give
the total corrected $B_{T}^{o}$ magnitude and $(B-V)_{T}^{o}$
colour index in the RC3 system. Log$R_{25}$ and the galactic
absorption $A_{B}$ were taken from RC3 Catalogue and
Burstein \& Heiles (\cite{bursteinheiles82}), respectively.
As stated above, total typical uncertainties in our photometry
are 0.15, 0.14, 0.15 and 0.14 for $B$, $V$, $R$ and $I$ bands.

\begin{table*}
  \caption[]{Magnitudes and Colour Indices}
\label{tabmagcolours}
\begin{tabular}{cccccccccc}
\hline\noalign{\smallskip}
KPG & $B$ & $V$ & $R$ & $I$ & $B-V$ & $B-I$& $B_{T}^{o}$ & $(B-V)_{T}^{o}$& 
Notes\\
\hline\noalign{\smallskip}

KPG64A  & 13.70 & 12.92 & 12.37 & 11.70 & 0.78 &  2.00 & 13.32 & 0.66& \\
KPG64B  & 15.17 & 14.21 & 13.66 & 12.98 & 0.96 &  2.18 & 14.44 & 0.75& \\
KPG68A  & 13.56 & 12.82 & 12.19 & 11.51 & 0.73 &  2.05 & 13.02 & 0.58& CP\\
KPG68B  & 14.73 & 13.98 & 13.31 & 12.65 & 0.75 &  2.08 & 14.07 & 0.58& CP\\
KPG75A  & 14.62 & 13.68 & 13.19 & 12.18 & 0.95 &  2.44 & 14.07 & 0.80& \\
KPG75B  & 15.36 & 14.49 & 13.99 & 12.97 & 0.87 &  2.39 & 14.50 & 0.65& \\
KPG88A  & 15.06 & 13.69 & 12.90 & 12.17 & 1.37 &  2.89 & 14.13 & 1.15& \\
KPG88B  & 16.12 & 14.85 & 14.11 & 13.44 & 1.27 &  2.68 & 15.21 & 1.05& \\
KPG98A  & 15.62 & 15.34 & 14.08 & 13.28 & 0.28 &  2.34 & 14.76 & 0.10& \\
KPG98B  & 15.51 & 14.22 & 13.03 & 12.07 & 1.29 &  3.44 & 14.89 & 1.12& \\
KPG102A & 15.68 & 15.18 & 14.84 & 13.87 & 0.51 &  1.81 & 14.98 & 0.29& \\
KPG102B & 15.06 & 14.57 & 14.22 & 13.42 & 0.49 &  1.64 & 14.41 & 0.29& \\
KPG103A & 16.27 & 15.48 & 14.99 & 14.18 & 0.79 &  2.10 & 15.66 & 0.61& \\
KPG103B & 15.12 & 14.47 & 14.11 & 13.42 & 0.64 &  1.69 & 14.62 & 0.47& \\
KPG108A & 15.17 & 13.95 & 13.49 & 12.55 & 1.22 &  2.62 & 14.31 & 1.01& \\
KPG108B & 14.35 & 13.17 & 12.67 & 11.76 & 1.17 &  2.59 & 13.45 & 0.95& \\
KPG112A & 13.24 & 12.41 & 11.85 & 11.19 & 0.83 &  2.05 & 12.33 & 0.60& CP\\
KPG112B & 13.47 & 12.63 & 12.05 & 11.35 & 0.84 &  2.12 & 12.85 & 0.68& CP\\
KPG112s & 12.55 & 11.78 & 11.20 & 10.52 & 0.77 &  2.03 &       &     & \\
KPG125A & 13.65 & 12.88 & 12.21 & 11.39 & 0.77 &  2.26 & 13.16 & 0.63& \\
KPG125B & 12.92 & 12.15 & 11.56 & 10.81 & 0.77 &  2.11 & 12.42 & 0.63& \\
KPG136A & 14.98 & 14.29 & 13.72 & 13.31 & 0.69 &  1.67 & 14.73 & 0.58& \\
KPG136B & 15.21 & 14.34 & 13.73 & 13.24 & 0.87 &  1.97 & 14.87 & 0.75& \\
KPG141A & 15.21 & 13.96 & 13.40 & 12.65 & 1.25 &  2.56 & 14.49 & 1.06& \\
KPG141B & 14.97 & 13.92 & 13.50 & 12.67 & 1.05 &  2.30 & 14.62 & 0.95& \\
KPG150A & 14.24 & 13.26 & 12.46 & 11.90 & 0.98 &  2.34 & 13.84 & 0.86& BS\\
KPG150B & 13.28 & 12.23 & 11.57 & 10.85 & 1.05 &  2.43 & 12.80 & 0.95& BS\\
KPG151A & 15.40 & 14.17 & 13.42 & 12.77 & 1.22 &  2.63 & 14.32 & 0.93& CP\\
KPG151B & 15.07 & 13.87 & 13.13 & 12.53 & 1.20 &  2.53 & 14.60 & 1.05& CP\\
KPG151s & 14.67 & 13.33 & 12.55 & 11.94 & 1.34 &  2.73 &       &     & \\
KPG156A & 13.02 & 12.60 & 12.11 & 11.71 & 0.42 &  1.31 & 12.61 & 0.30& CP\\
KPG156B & 14.54 & 13.93 & 13.42 & 12.95 & 0.61 &  1.59 & 14.26 & 0.52& CP\\
KPG159A & 16.39 & 15.89 & 15.39 & 15.45 & 0.50 &  0.94 & 15.90 & 0.36& \\
KPG159B & 14.61 & 13.88 & 13.34 & 12.95 & 0.73 &  1.66 & 14.16 & 0.60& \\
KPG160A & 13.98 & 13.09 & 12.58 & 11.74 & 0.89 &  2.23 & 13.80 & 0.83& \\
KPG160B & 15.44 & 14.57 & 14.07 & 13.23 & 0.87 &  2.22 & 14.83 & 0.71& \\
KPG168A & 12.80 & 11.82 & 11.21 & 10.68 & 0.99 &  2.12 & 12.29 & 0.86& CP\\
KPG168B & 15.16 & 14.23 & 13.73 & 13.25 & 0.93 &  1.91 & 14.49 & 0.77& CP\\
KPG195A & 13.24 & 12.38 & 11.81 & 11.20 & 0.87 &  2.04 & 12.87 & 0.77& \\
KPG195B & 14.42 & 13.71 & 13.28 & 12.78 & 0.71 &  1.64 & 13.88 & 0.58& \\
KPG211A & 13.68 & 12.69 & 12.01 & 11.43 & 0.99 &  2.25 & 13.45 & 0.91& \\
KPG211B & 15.75 & 14.70 & 13.99 & 13.39 & 1.05 &  2.36 & 15.05 & 0.87& \\
KPG216A &       &       &       &       &      &       &       &     & BS\\
KPG216B & 13.28 & 12.70 & 12.37 & 11.67 & 0.58 &  1.62 & 12.88 & 0.36& BS\\
KPG249A & 12.45 & 12.10 & 11.67 & 11.42 & 0.35 &  1.03 & 12.26 & 0.30& CP\\
KPG249B & 12.93 & 12.50 & 12.01 & 11.68 & 0.43 &  1.24 & 12.51 & 0.32& CP\\
KPG249s & 11.89 & 11.49 & 11.03 & 10.75 & 0.40 &  1.14 &       &     & \\
KPG295A & 13.45 & 12.59 & 11.97 & 11.42 & 0.86 &  2.03 & 13.23 & 0.79& CP\\
KPG295B & 13.30 & 12.50 & 11.92 & 11.34 & 0.80 &  1.97 & 12.86 & 0.69& CP\\
KPG302A & 11.23 & 10.67 & 10.28 &  9.64 & 0.56 &  1.59 & 11.03 & 0.51& \\
KPG302B & 14.05 & 13.57 & 13.18 & 12.56 & 0.46 &  1.47 & 13.90 & 0.42& \\
KPG313A & 13.17 & 12.52 & 12.10 & 11.87 & 0.64 &  1.30 & 13.09 & 0.62& \\
KPG313B & 13.26 & 12.17 & 11.39 & 10.68 & 1.09 &  2.57 & 12.97 & 1.02& \\
KPG332A & 12.19 & 11.35 & 10.97 & 10.14 & 0.84 &  2.05 & 11.91 & 0.77& \\
KPG332B & 12.71 & 11.61 & 11.11 & 10.09 & 1.10 &  2.62 & 12.04 & 0.94& \\
\hline\noalign{\smallskip}
\end{tabular}
\\CP = Pair apparently in Contact \\
BS = Bright Star nearby in the Field \\
\end{table*}

\setcounter{table}{2}

\begin{table*}
  \caption[]{Continued.}
\begin{tabular}{cccccccccc}
\hline\noalign{\smallskip}
KPG & $B$ & $V$ & $R$ & $I$ & $B-V$ & $B-I$& $B_{T}^{o}$ & $(B-V)_{T}^{o}$\\
\hline\noalign{\smallskip}
KPG347A & 12.13 & 11.34 & 10.85 & 10.26 & 0.79 &  1.87 & 11.97 & 0.74& CP\\
KPG347B & 12.11 & 11.19 & 10.61 &  9.96 & 0.92 &  2.15 & 11.80 & 0.83& CP\\
KPG347s & 11.40 & 10.52 &  9.96 &  9.39 & 0.89 &  2.02 &       &     & \\
KPG389A & 13.69 & 12.99 & 12.37 & 12.15 & 0.71 &  1.55 & 13.40 & 0.61& CP\\
KPG389B & 13.67 & 12.83 & 12.23 & 11.62 & 0.84 &  2.06 & 13.47 & 0.76& CP\\
KPG396A & 14.78 & 14.26 & 13.84 & 13.17 & 0.52 &  1.61 & 14.18 & 0.35& \\
KPG396B & 14.06 & 13.65 & 13.28 & 12.48 & 0.41 &  1.57 & 13.99 & 0.36& \\
KPG404A & 13.84 & 13.12 & 12.60 & 12.08 & 0.72 &  1.76 & 13.61 & 0.65& CP\\
KPG404B & 12.57 & 11.81 & 11.26 & 10.63 & 0.76 &  1.93 & 12.33 & 0.68& CP\\
KPG426A & 14.77 & 13.78 & 13.14 & 12.57 & 0.99 &  2.20 & 14.46 & 0.88& \\
KPG426B & 15.01 & 14.11 & 13.50 & 13.02 & 0.90 &  1.99 & 14.84 & 0.82& \\
KPG440A & 13.22 & 12.94 & 12.63 & 12.81 & 0.28 &  0.41 & 13.01 & 0.22& \\
KPG440B & 12.42 & 11.34 & 10.87 &  9.90 & 1.08 &  2.52 & 11.79 & 0.92& \\
KPG455A & 13.99 & 13.11 & 12.61 & 11.80 & 0.88 &  2.19 & 13.66 & 0.78& \\
KPG455B & 13.10 & 12.47 & 12.02 & 11.17 & 0.63 &  1.93 & 12.61 & 0.50& \\
\hline\noalign{\smallskip}
\end{tabular}
\\CP = Pair apparently in Contact
\end{table*}

Magnitude and colour corrections were not applied for 
a few small galaxies (blank spaces in Table) in the neighborhood 
of our pairs, due to a lack of reliable information.
Our observations span a range (11, 15.9) and (0.3, 1.1) mag
in $B_{T}^{0}$ and $(B-V)_{T}^{0}$, respectively. The
observed $(B-V)$ range is comparable (by judging the
colour maps scales) to that in a similarly selected sample 
of pairs in the southern hemisphere by Reduzzi \& Rampazzo
 (\cite{reduzzirampazzo96}), although some E/S0 components were 
included in that sample. Interestingly, our $(B-V)_{T}^{0}$ 
range is comparable to the full range found in Larson \& Tinsley 
(\cite{larsontinsley78}) in spite of the fact that their 
interacting sample is biased in favour of strongly peculiar 
systems from the Arp's catalogue. Similarly, the photoelectric 
Cousins $UBVRI$ photometry of interacting galaxies by Johansson 
\& Bergvall (\cite{johanssonbergvall90}) shows a 
comparable range in the observed $(B-V)$ colours, although 
this sample is biased in favour of disturbed morphology,
the presence of bridges and includes a fraction of E/S0 
components.

\subsection{The Holmberg Effect}

\begin{figure}
%  \resizebox{\hsize}{!}{\includegraphics{bv1_bv2.eps}}
\resizebox{\hsize}{!}{\includegraphics{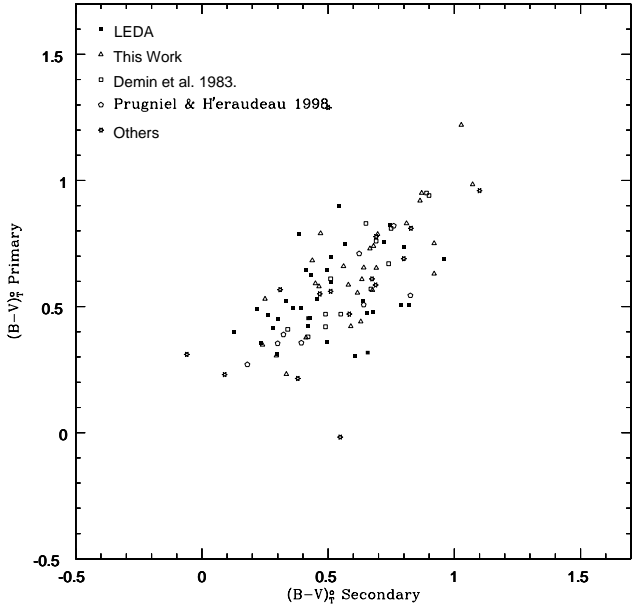}}
\caption{The Holmberg effect. $(B-V)^{0}_{T}$ primary versus 
$(B-V)^{0}_{T}$ 
secondary.}
\label{fig_holmberg}
\end{figure}

As a byproduct of his famous photometric survey of nearby 
galaxies, Holmberg (\cite{holmberg58}) compared the
photographic colours of paired galaxies and found a significant 
correlation between the colours of pair components. This 
phenomenon has since been referred to as the ``Holmberg effect''.
Figure \ref{fig_holmberg} shows the correlation between the 
$(B-V)_{T}^{o}$ colour index. In a few cases irregular 
galaxies belonging to pairs have conventionally been considered 
as spirals. The colour index along the vertical axis refers to 
the brighter (primary) component and that along the horizontal 
axis refers to the fainter (secondary) component in each pair. 
To reinforce the validity of any correlation, other symbols 
indicate sources of $(B-V)_{T}^{o}$ data for additional (S+S)
Karachentsev pairs from the literature.

The colour correlation between pair components
is tight. A correlation coefficient $r \sim 0.77$ with a
residual sigma of 0.18 is obtained for our (S+S) data.
Additionally, a fitting to all the data in Figure \ref{fig_holmberg},
give a correlation coefficient $r \sim 0.70$ with a residual sigma 
of 0.18. All the (S+S) pairs with ($B-V$) either 
from our observations or from the literature, have a median 
relative velocity, $\Delta v \sim 45$ km s$^{-1}$ and a median 
projected separation $\Delta x \sim$ 29 kpc.
Although the physical explanation of the Holmberg effect 
is complex, it has been interpreted as reflecting a 
tendency for similar types of galaxies to form together 
(morphological concordance), a possible reflection of the 
role of local environment in determining galaxy morphology, 
but alternatively, it can presumably also reflect mutually 
induced star formation (Kennicutt et al. 
\cite{kennicuttetal87}) in physical pairs.

\subsection{Colours and Morphology} 

\begin{figure*}
%\resizebox{\hsize}{!}{\includegraphics{}}
\includegraphics{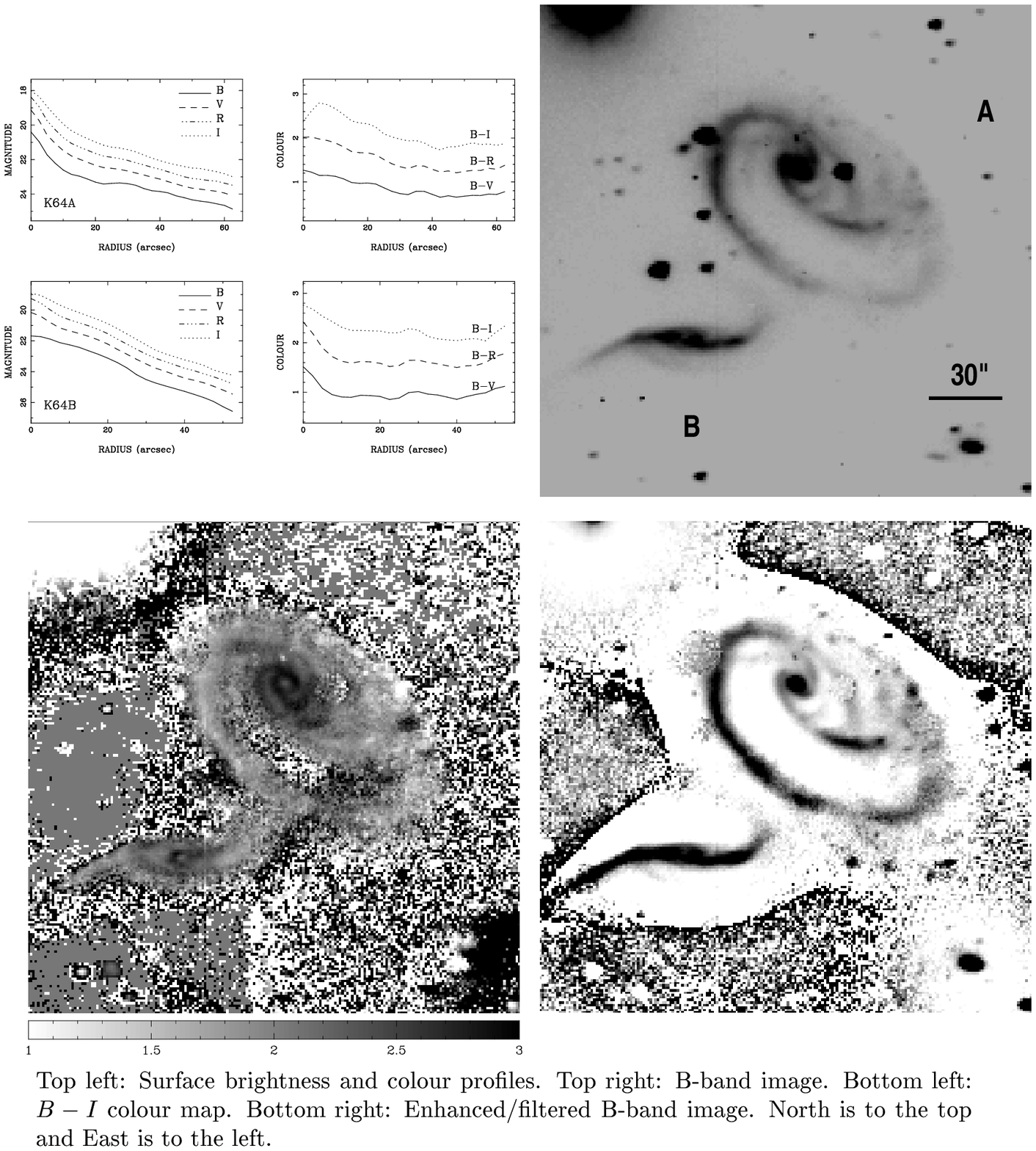}
\vskip600pt
\caption{KPG64 Mosaic.}
\label{fig_mosaics}
\end{figure*}

To discuss the optical morphology (that could be 
modified by the presence of bars, spiral arms, rings, etc)
and its relationship to the global photometrical properties, 
the final results for each pair are presented in the form 
of a mosaic (Figure \ref{fig_mosaics}) including: 1) mean
azimuthally averaged surface brightness and colour profiles, 
2) gray scale $B$-band images, 3) ($B-I$) colour 
index maps and 4) $B$ band-filtered images for each pair. 
In most of the cases, not all the foreground stars in each 
field have been removed. The images in the lower panels 3) 
and 4) can be combined to look for morphological features 
like the presence of bars, rings, the shape of spiral arms, 
the presence of tidal features and other morphological 
distortions presumably associated with the interactions. 
The filtered/enhancing techniques (Sofue \cite{sofue93})
applied in 4), allow the subtraction of the diffuse
background in a convenient way to discuss different
morphological details.

Karachentsev (\cite{kara72}) identified three 
basic interaction classes (AT, LI and DI) that describe 
the pairs which show obvious signs of interaction. AT 
class identifies pairs with components in a common 
luminous halo with a symmetric, amorphous or shredded, 
asymmetric (sh) structure. LI pairs show evidence of 
tidal bridges (br), tails (ta) or both (br+ta). DI pairs 
show evidence of structural distortion in one (1) or both 
(2) components. We add to this sequence NI for the (S+S) 
pairs with no obvious morphological distortion. The order 
AT-LI-DI-NI can be regarded as a sequence from strongest 
to weakest evidence for tidal distortion or, alternatively, 
most to least dynamically evolved (interpreting a common 
envelope as a sign of extensive dynamical evolution in 
pairs).

Based on our CCD observations, it is now possible  
to attempt 1) a reclassification of the Hubble 
morphology from a combination of our colour index $(B-I)$ 
maps and sharp/filtered $B$ images, 2) a reclassification 
of the global pair interaction morphology (I/A class hereafter) 
in the Karachentsev Catalogue and 3) a reclassification of the 
spiral arm morphology as suggested by Elmegreen \& Elmegreen 
(\cite{elmegreenelmegreen82}) (hereafter EE class).

It is known that the colours of spiral galaxies are 
correlated with its morphological type T. Although the 
colour indices of galaxies belonging to type T will have a 
large dispersion, the median value declines systematically 
as T increases along the morphological sequence. Median 
integrated total $(B-V)$ colours of galaxies according to 
morphological class are given by Roberts \& Haynes 
(\cite{robertshaynes94}). The UGC and the Local 
Supercluster (LSc) samples in Roberts \& Haynes 
(\cite{robertshaynes94}) are rather inhomogeneous in 
terms of environment, but the interacting objects were 
excluded from their analysis. We may consider these samples
as comparison/reference samples for the following 
discussion.

\subsection{Comments on Individual Objects}

Before proceeding to a discussion of the results shown
in tables and figures, we present comments on morphological
details found in individual pairs that may be relevant to 
our conclusions and look for any relationship to the global 
photometrical profiles. 
\medskip

{\bf KPG64A}. The galaxy is classified as SA(s)b pec. 
Our $(B-I)$ colour map shows two centrally symmetric spiral 
arms that become bifurcated at the outer parts. The arms 
are sharp-defined and the B-band sharp/filtered image can 
show some well define knotty structures along them. A bright 
small nucleus can also be appreciated. We classify this
galaxy as Sc pec. The type of arms shown may probably be 
produced/modified by the interaction as predicted by 
simulations in Noguchi (\cite{noguchi90}). The total 
$(B-V)_{T}^{0}$ colour is representative of
Sab-Sb types. Our EE class is 6.

{\bf KPG64B}. The galaxy is classified as SB(s)a pec. 
Our $(B-I)$ colour map show an apparently inclined
galaxy that could be simulating a barred structure in the
central region of a lower resolution image. The outer spiral
structure resembles an integral sign that could be tidally
generated by the interaction, on line with simulations
by Noguchi (\cite{noguchi90}). The sharp/filtered image shows 
a faint bifurcated structure emanating from the eastern arm. 
There is also evidence of knotty structure along the arms, 
but no clear evidence of a bar structure. We classify this
galaxy as Sbc. The total $(B-V)_{T}^{0}$ colour is more
representative of S0a-Sa types. We notice that the colour
profiles tend to be flat after 15$\arcsec$. Our EE class
is 6. The I/A class for the pair is LI.

{\bf KPG68A}.  This galaxy is overlapping at the southeast
with its companion. Our iterative modeling of the light 
distribution is poor and caution is needed with the total 
magnitudes and colours for both galaxies. The galaxy is 
classified as a Scd:. Our sharp/filtered and $B-I$ colour 
map images shows a bright nucleus plus multiple arms with 
knotty structure along them. The total $(B-V)_{T}^{0}$ 
colour is representative of Sb-Sbc types. Our EE class
is 3.

{\bf KPG68B}. This galaxy is overlapping at the north
with its companion. The galaxy is classified as a SBb:. Our
sharp/filtered and $B-I$ colour map images shows knotty structure 
along the arms from either end of a prominent bar. The total 
$(B-V)_{T}^{0}$ colour is representative of Sb-Sbc types.
The colour profiles in this barred galaxy are flat after 
5$\arcsec$ from its center. The I/A class for the pair is 
DI. Our EE class is 10.

{\bf KPG75A}.  The galaxy is classified as E?, but our $(B-I)$
and sharp/filtered images show a bright central nucleus from
which two opened and diffuse spiral arms emanate. The spiral 
pattern shows an integral sign (perhaps as a by product of 
the interaction). We classify this galaxy as Sab. The 
$(B-V)_{T}^{0}$ colour is representative of S0a-Sa types.

{\bf KPG75B}.  The galaxy is classified as SB?, and our $(B-I)$
and sharp/filtered images show a bright nucleus and two 
wrapped adjacent spiral arms that in projection may
simulate a bar. We can not clearly appreciate a bar 
structure. The arm at the north appears warped.
We classify this galaxy as Sb. The $(B-V)_{T}^{0}$ 
colour is representative of Sab-Sb types. The I/A class
for the pair is DI. Our EE class is 7.

{\bf KPG88A}. The galaxy is classified as a SA(s)c 
and our $(B-I)$ and sharp/filtered images shows a few 
bright knots along multiple arms that emanate from a 
bright nucleus. The $(B-V)_{T}^{0}$ colour is 
redder than that corresponding to its 
morphological type. Our estimated $(B-V)$ colour is 
consistent, however, with that reported by Prugniel 
\& H\'eraudeau (\cite{prugnielheraudeau98}). The colour
profiles tend to be flat after 15$\arcsec$. Our EE
class is 12.

{\bf KPG88B}. The galaxy is classified as a SBcd:. Both our
sharp/filtered and $(B-I)$ colour map show a prominent bar
and complex arms extending from either end. The 
$(B-V)_{T}^{0}$ colour is redder than that corresponding to 
its morphological type. The I/A class for the pair is DI. 
Our EE class is 10.

{\bf KPG98A}.  The galaxy is classified as Scd:. Our $(B-I)$ and
sharp/filtered images show a bright nucleus plus a perturbed
spiral pattern that seems warped in projection. A few knots
can be appreciated along the arms. The $(B-V)_{T}^{0}$ colour 
is bluer than that corresponding to its morphological 
type. We caution the reader about an aparent inconsistency
in the results obtained for this pair. Contrary to the 
observed $(B-V)$ value, the corresponding colour profiles 
show a tendency to be redder. After repeating the calculations 
and taking into account our estimated errors in B and V 
magnitudes, we do not have an explantion for this behaviour.
Our EE class is 2.

{\bf KPG98B}.  This galaxy is classified as S?. Our $(B-I)$
and sharp/filtered images show a prominent nucleus plus a faint
but defined spiral pattern that simulates an outer 
pseudo-ring structure. This pattern could be reminiscent of
the interaction with its companion. We classify this galaxy
as (R)Sa. The $(B-V)_{T}^{0}$ colour is redder than
that corresponding to its morphological type. RC3 Catalogue
reports (only) a blue photographic magnitude with an error 
of 0.2 mag that was transformed to the standard system. 
However, this value is within 0.2 mag to ours.
The I/A class for the pair is DI.

{\bf KPG102A}. This galaxy is classified as Sa. Our $(B-I)$ 
and sharp/filtered images show a prominent nuclear
structure from which a diffuse wrapped spiral 
pattern emerges. We classify the galaxy as SBab. The 
$(B-V)_{T}^{0}$ colour is bluer than that corresponding
to its morphological type.

{\bf KPG102B}. The galaxy is classified as Sb and our 
$(B-I)$ and sharp/filtered images show a bright  
nuclear region plus a beautiful symmetric and 
sharp-defined spiral pattern that could be reminiscent 
of the interaction with its companion. We classify the
galaxy as Sc. The $(B-V)_{T}^{0}$ colour is bluer than
that corresponding to its morphological type. The 
$(B-R)$ and $(B-I)$ colour profiles have a tendency to
be flat after 10$\arcsec$. The I/A class for the pair 
is LI. Our EE class is 11.

{\bf KPG103A}. The galaxy is classified as Sb. Its 
small angular size make the visualization of features a 
difficult task. At our resolution, both the $(B-I)$ and 
sharp/filtered images shows a peculiar morphology. Keel
(\cite{keel88}) has extensively studied this galaxy. The 
$(B-V)_{T}^{0}$ colour is representative of a Sab-Sb 
galaxy. The $(B-V)$ colour profile have a tendency to be
flat along the observed radius.

{\bf KPG103B}. The galaxy is classified as Sa. 
At our resolution, the $(B-I)$, sharp/filtered
and unsharp masking images show a peculiar morphology. 
The arms appear radially distributed from a prominent 
bulge. Keel (\cite{keel88}) has extensively studied 
this galaxy. The $(B-V)_{T}^{0}$ colour is representative 
of a Scd-Sd galaxy. The I/A class for the pair is DI.

{\bf KPG108A}.  The galaxy is classified as Sbc. Our $(B-I)$ 
and sharp/filtered images show a highly inclined galaxy with 
a bright elongated bulge and a complex dusty structure. 
It is difficult to find definite signs of perturbation.
The $(B-V)_{T}^{0}$ colour is redder than that
corresponding to its morphological type. RC3 Catalogue
reports (only) a blue photographic magnitude with an error 
of 0.2 mag that was transformed to the standard system. 
However, this value is within 0.2 mag to ours.
The $(B-R)$ and $(B-V)$ colour profiles have a tendency to 
be flat after 25$\arcsec$.

{\bf KPG108B}.  The galaxy is classified as Sb. Our $(B-I)$ 
and sharp/filtered images show a highly inclined galaxy with 
a complex dusty structure. It is difficult to trace 
signs of perturbation. A small galaxy (north-east) in its 
neighborhood can be appreciated. The $(B-V)_{T}^{0}$ colour 
is redder than that corresponding to its morphological type. 
RC3 Catalogue reports (only) a blue photographic magnitude with 
an error of 0.2 mag that was transformed to the standard 
system. However, this value is within 0.2 mag to ours.
The colour profiles have a tendency to be flat after 
30$\arcsec$. The I/A class for the pair is NI.

{\bf KPG112A}.  The galaxy is classified as S0/a.
Our sharp/filtered and $B-I$ images show and edge-on
galaxy resembling a lenticular or an early-type spiral
with a distorted disk. The distorted disk may be representing 
a tidal tail/counter-tail structure generated by the interaction.
We classify this galaxy as Sa. The $(B-V)_{T}^{0}$ colour is 
representative of Sab-Sb types. The $(B-V)$ and $(B-R)$ colour 
profiles tend to be flat after 10$\arcsec$, while the $(B-I)$ 
colour profile appear flat all along the observed radius.

{\bf KPG112B}.  The galaxy is classified as S0:. Our
sharp/filtered and $B-I$ images show a bright prominent  
bulge. Two diffuse spiral arms appear wrapped. The arms 
at the west side are seen, in projection, overlapping at
the eastern arm of its companion galaxy. We classify this
galaxy as Sa. The $(B-V)_{T}^{0}$ colour is 
representative of a Sa galaxy. Our EE class is 12.
The I/A class for the pair is LI.

{\bf KPG125A}. The galaxy is classified as Pec. In spite 
of its small angular size, our $(B-I)$ and sharp/filtered 
images show two faint spiral arms emerging from a complex and
bright central region. We classify this galaxy as Sab pec. 
The estimated $(B-V)_{T}^{0}$ colour is representative of 
Sab-Sb types. The colour profiles in this galaxy show a  
tendency to be flat after 5$\arcsec$.

{\bf KPG125B}. The galaxy is classified as S pec. However, 
both our $(B-I)$ and sharp/filtered images show multiple 
and complex spiral arms with knotty features along them and
emanating from a bright nuclear region . We classify
this galaxy as Sc pec. The estimated $(B-V)_{T}^{0}$ colour
is representative of Sab-Sb types. The colour profiles  
show a tendency to be flat after 20$\arcsec$. The 
I/A class for the pair is DI. Our EE class is 9.

{\bf KPG136A}.  This galaxy is classified as S?. Our $(B-I)$ and
sharp/filtered images show a bright central region from which 
multiple but diffuse arms appear to emanate. We classify this
galaxy as Sbc. The estimated $(B-V)_{T}^{0}$ colour is 
representative of Sbc-Sc types. Our EE class is 10.

{\bf KPG136B}.  This galaxy is classified as S?. Our $(B-I)$ and
sharp/filtered images show a bright nuclear region surrounded by
a tightly wrapped arm-like structure resembling a ring or
pseudo-ring. An outer faint feature also resembles a diffuse 
shell/arc that may be associated with a tidal origin. We classify
the galaxy as S(r)ab. The $(B-V)_{T}^{0}$ colour is representative 
of a Sa type. The I/A class for the pair is DI. Our EE class is 8.

{\bf KPG141A}.  The galaxy is classified as S?. Our $(B-I)$ 
and sharp/filtered images show a highly inclined galaxy where 
a central bulge and a few knots along a thin linear feature 
(arm seen in projection?) at the north-east can be appreciated. 
We classify this galaxy as Sbc. The $(B-V)_{T}^{0}$ colour is 
redder than that representative of its morphological 
type. RC3 Catalogue reports (only) a blue photographic magnitude 
with an error of 0.3 mag that was transformed to the standard 
system. However, this value is within 1.1 mag to ours.

{\bf KPG141B}.  The galaxy is classified as S?. Our $(B-I)$ and 
sharp/filtered images show a bright prominent central region
and two wrapped but defined arms. We classify this galaxy as Sb.
The estimated  $(B-V)_{T}^{0}$ colour is redder than 
that representative of its morphological type. RC3 Catalogue
reports (only) a blue photographic magnitude with an error 
of 0.3 mag that was transformed to the standard system. 
However, this value is within 0.3 mag to ours. The I/A class
for the pair is DI. Our EE class is 7.

{\bf KPG150A}.   The galaxy is classified as Sa. Our $(B-I)$ 
and sharp/filtered images show an internal two-arm spiral 
pattern and a bright nucleus. An outer spiral arm pattern is 
wrapped and may be resembling, in projection, a pseudo-ring. 
We classify this galaxy as SA(r)b. The estimated $(B-V)_{T}^{0}$
colour is representative of S0-S0a types. Our EE class is 7.

{\bf KPG150B}. The galaxy is classified as SBb. 
Our $(B-I)$ and sharp/filtered images show a sharply defined 
bar and multiple knotty arms wrapped enough in the central
region to resemble an internal ring. We classify this galaxy as
SB(r)c. The colour profiles (in the presence of a bright nearby 
field star) do not show a tendency to be flat like other 
barred galaxies in this sample. The estimated $(B-V)_{T}^{0}$ 
colour is redder than that representative of its morphological
type. Prugniel \& H\'eraudeau (1998) report a blue magnitude
in agreement (for a similar aperture) to ours. RC3 Catalogue
reports (only) a blue total magnitude with an error of 0.15 
mag that was transformed to the standard system. However, 
this value is within 0.2 mag to ours. The I/A class for the 
pair is NI. Our EE class is 8.

{\bf KPG151A}.  The galaxy is seen slightly overlapping,
in projection, to its companion galaxy at the south-east
The galaxy is classified as Sc. Our $(B-I)$ and sharp/filtered 
images show an edge-on galaxy with a clearly defined bulge
region. The estimated $(B-V)_{T}^{0}$ colour is redder
than that corresponding to its morphological type. We 
could not find a reference in the literature to compare 
magnitudes and colours for this source.

{\bf KPG151B}. The galaxy is classified as SB?. Our $(B-I)$ 
and sharp/filtered images show a bright prominent nuclear
region and an adjacent elongated feature that resembles
a bar. We classify this galaxy as SBb. The $(B-V)_{T}^{0}$ 
colour is redder than that corresponding to its 
morphological type. RC3 Catalogue reports (only) a blue 
photographic magnitude with an error of 0.2 mag that was 
transformed to the standard system. However, this value 
is within 0.1 mag to ours. The $(B-V)$ colour profile shows a 
marginal tendency to be flat after 20$\arcsec$. The I/A class 
for the pair is LI. Our EE class is 8.

{\bf KPG156A}. The galaxy is classified as SA(r)c pec. Our 
sharp/filtered and $(B-I)$ images show a bright central nucleus 
surrounded by two knotty arms forming an inner ring structure. 
The arms extend far from the center forming: 1) a bridge to its 
companion and 2) a very long tail. They may be tidally-generated 
by the interaction. In addition, faint filamentary structures 
are seen almost tangent to the ring. The estimated $(B-V)_{T}^{0}$ 
colour is definitely bluer than that corresponding to its
morphological type. Our EE class is 11.

{\bf KPG156B}. The galaxy is classified as SB(rs)c pec. 
However, in our sharp/filtered and $(B-I)$ images no clear
ringed structure is appreciated. We notice instead, 
a bright and somehow elongated nuclear region with two 
diffuse and opened spiral arms resembling a integral sign. 
We classify this galaxy as SBbc pec. The estimated $(B-V)_{T}^{0}$ 
colour is representative of Sbc-Sc types. The $(B-R)$ and
$(B-I)$ colour profiles show a tendency to be flat after
15$\arcsec$. The I/A class for the pair is LI.

{\bf KPG159A}. The galaxy is classified as Sb. The small
angular size of this galaxy does not allow both the 
sharp/filtered and $(B-I)$ images to show any detailed 
morphology. The $(B-V)_{T}^{0}$ colour is representative of 
Sm-Im types.

{\bf KPG159B}. The galaxy is classified as Sb. The sharp/filtered
and $(B-I)$ images shows an inclined galaxy with a bright 
nuclear region and wrapped spiral arms. A faint linear feature 
crossing the central region resembles a bar. We classify this 
galaxy as SBb. The $(B-V)_{T}^{0}$ colour is representative 
of Sb-Sbc types. The I/A class for the pair is DI.

{\bf KPG160A}. The galaxy is classified as SB(s)a. Our 
$(B-I)$ and sharp/filtered images show a bright outer
ring enclosing a bar-like feature. The outer ring is 
bluer than the adjacent disk. We classify this galaxy as 
(R')SB(s)a. The $(B-V)_{T}^{0}$ colour is representative 
of S0a-Sa types. The colour profiles show a tendency to
be flat after 25$\arcsec$. Our EE class is 8.

{\bf KPG160B}.  The galaxy is classified as SBa. Our  
$(B-I)$ and sharp/filtered images show a highly 
inclined system where it is difficult to see the bar 
and bulge regions. There is knotty structure along the 
main body of the galaxy. We classify this galaxy as Sb. 
The $(B-V)_{T}^{0}$ colour is representative of Sa-Sab 
types. If the outer ring in the companion galaxy is regarded 
as evidence of interaction, the I/A class for the pair is DI, 
otherwise is NI.

{\bf KPG168A}. The galaxy is classified as Sa. 
Our $(B-I)$ and sharp/filtered images show a prominent bulge 
region and two symmetric spiral arms that simulate a 
pseudo-ring. In the external parts, the arms are extended 
and diffuse (resembling an integral sign) forming a bridge at 
the south-east to its companion galaxy. We classify this galaxy 
as S(s)b. The $(B-V)_{T}^{0}$ colour is more representative of 
S0-S0a types. The colour profiles show a tendency to be flat 
from  40$\arcsec$. Our EE class is 6.

{\bf KPG168B}. The galaxy is classified as Sc. Our $(B-I)$ and 
sharp/filtered images show an apparently inclined system with a 
few prominent knots along the main body. Two adjacent diffuse arms 
are also appreciated. One of them is apparently forming a
bridge at the west to its companion galaxy. The $(B-V)_{T}^{0}$ 
colour is more representative of Sa-Sab types. The colour profiles 
show a tendency to be flat along most of the observed radius. 
The I/A class for the pair is LI.

{\bf KPG195A}. The galaxy is classified as SB(s)a pec. 
The configuration of this pair resembles that of KPG168. Our 
$(B-I)$ and sharp/filtered images show a prominent bulge 
and an adjacent linear feature that crosses the central region
resembling a bar. From this bar, two spiral arms emerge. These 
arms are prominent in the central regions and become diffuse 
and extended (resembling an integral sign) at the external 
parts. We classify this galaxy as SB(s)b pec. The $(B-V)_{T}^{0}$ 
colour is representative of S0a-Sa types The colour profiles 
show a marginal tendency to be flat after 40$\arcsec$. Our 
EE class is 6.

{\bf KPG195B}. The galaxy is classified as SB(s)m? Our 
$(B-I)$ and sharp/filtered images show an apparently inclined 
system with bright condensations along the main body and two 
adjacent arms that become diffuse at the outskirt.  
The bar structure is difficult to appreciate. We classify this
galaxy as Sc. The $(B-V)_{T}^{0}$ colour is representative of 
Sbc-Sc types. The I/A class for the pair is DI.

{\bf KPG211A}. The galaxy is classified as (R')SAB(rs)ab pec: 
and our $(B-I)$ and sharp/filtered images seem to confirm this 
classification. The pattern of spiral arms is complex, tightly 
wrapped and show blue colours. This could be a sign of strong 
perturbation from its companion. The estimated $(B-V)_{T}^{0}$ is 
representative of S0-S0a types. The colour profiles show a 
tendency to be flat after 30$\arcsec$. Our EE class is 8.

{\bf KPG211B}. The galaxy is classified as Sb: and our $(B-I)$ 
and sharp/filtered images show an apparently inclined galaxy 
with a prominent bulge and two symmetric, diffuse spiral arms.
We classify this galaxy as Sa. The $(B-V)_{T}^{0}$ colour is 
representative of S0-S0a types. Our EE class is 7. The I/A class 
for the pair is DI.

{\bf KPG216A}. The galaxy has a bright nearby field star that 
was difficult to subtract in our iterative procedure. The galaxy 
is classified as SB(s)b pec: Our $(B-I)$ and sharp/filtered 
images show an elongated central region resembling a bar structure 
from which two diffuse spiral arms (integral sign) emerge.
Our EE class is 10.

{\bf KPG216B}. The galaxy is classified as SAB(s)c pec:. Our 
$(B-I)$ and sharp/filtered images show an elongated feature
crossing the center and resembling a bar from which two 
knotty arms emanate. The arm at the west is multiple.
We classify this galaxy as SBc pec. The $(B-V)_{T}^{0}$ colour 
is representative of Sm-Im types. The colour profiles show 
structure and a global tendency to be flat after 35$\arcsec$. 
The I/A class for the pair is DI. Our EE class is 9.

{\bf KPG249A}. The pair show an apparent low degree of overlapping. 
The galaxy is classified as SAB(rs)cd pec. Our sharp/filtered 
and $(B-I)$ images show two bright condensations in the nuclear 
region. The arms show a bifurcated spiral pattern and knotty 
features. The arm at north-east is forming an apparent bridge to 
its companion galaxy. We classify this galaxy as SABcd pec. The  
$(B-V)_{T}^{0}$ colour is more representative of Sm-Im types.  
The colour profiles show a tendency to be flat after 30$\arcsec$. 
Our EE class is 6.

{\bf KPG249B}. The galaxy is classified as IBm pec. Our
sharp/filtered and $(B-I)$ images show some bright 
condensations along a main elongated body that resembles
a bar structure from which two diffuse opened arms emerge. 
We classify this galaxy as SBm pec. The $(B-V)_{T}^{0}$ colour 
is more representative of Sm-Im types. The colour profiles show 
a tendency to be flat along the observed radius. The I/A class 
for the pair is LI.

{\bf KPG295A}. This is a low-degree overlapping pair where 
both components show remarkably similar morphological 
features. The galaxy is classified as SAB(rs)a pec and both 
our $(B-I)$ and sharp/filtered images show a bright nuclear 
region and a faint adjacent broad feature that may be resembling
a bar structure. These features are enclosed by an internal 
set of blue arms forming an elongated internal ring. 
In addition, this galaxy also shows an external diffuse and 
elongated ring. We classify this galaxy as (R')SAB(r)a pec. 
The $(B-V)_{T}^{0}$ colour is representative of S0a-Sa types.  
Our EE class is 8.

{\bf KPG295B}. The galaxy is classified as SAB(rs)ab pec. Our
$(B-I)$ and sharp/filtered images show a galaxy with a prominent
nuclear region enclosed by two wrapped arms resembling an inner 
ring structure. The external arms are bright, blue and 
wrapped. The $(B-V)_{T}^{0}$ colour is representative of 
Sab-Sb types. The I/A class for the pair is DI. Our EE class
is 8.

{\bf KPG302A}. The galaxy is classified as SAB(rs)c:. Our $(B-I)$ 
and sharp/filtered images show a beautiful spiral structure with 
knotty blue features all along the arms. We can not appreciate  
signs of a barred structure. We classify this galaxy as Sc. The
$(B-V)_{T}^{0}$ colour is representative of Sc-Scd types. Our EE 
class is 12.

{\bf KPG302B}. The galaxy is classified as SB0/a pec. Our 
$(B-I)$ and sharp/filtered show a prominent and elongated
central region that may resemble a bar. Alternatively,
we may interpret that as two overlapping bright central 
sources forming an elongated feature from which two 
diffuse spiral arms emanate. The arm towards the companion
is apparently bifurcated forming an external arc/shell-like 
feature. We classify this galaxy as SBbc pec. The $(B-V)_{T}^{0}$ 
colour is representative of Sd-Sm types. The colour profiles 
show a tendency to be flat all along its radius.
The I/A class for the pair is DI. Our EE class is 6.

{\bf KPG313A}. The galaxy is classified as 
SAB(rs)cd. Our $(B-I)$ and sharp/filtered images show a 
central elongated bar-like feature that is enclosed by a set 
of arms with apparently differing pitch angles and resembling 
a broken ring. Blue knotty features are appreciated along the 
arms. The $(B-V)_{T}^{0}$ colour is representative of Sb-Sbc 
types. The $(B-I)$ colour profile show a tendency to be flat 
after 40$\arcsec$. Our EE class is 1.

{\bf KPG313B}. The galaxy is classified as Sab:sp. Our
sharp/filtered and $(B-I)$ images show an inclined galaxy 
with extended and apparently warped arms. The $(B-V)_{T}^{0}$ 
colour is redder than that corresponding to its morphological 
type. RC3 reports total asymptotic $B$ and $V$ magnitudes 
with associated errors of 0.15 mag that imply a $(B-V)$ colour 
in agreement to our observed value. The I/A class for the pair 
is DI.

{\bf KPG332A}. The galaxy is classified as SA(rs)c and our $(B-I)$ 
and sharp/filtered images show a bright nucleus and blue knotty 
arms. We do not find evidence for an internal ring or s-shaped
structure. We classify this galaxy as Sc. The $(B-V)_{T}^{0}$ 
colour is representative of Sa type. Our EE class is 3.

{\bf KPG332B}. The galaxy is classified as Sc: sp 
and our $(B-I)$ and sharp/filtered images show a spectacular 
and complex dust lane structure all along the plane of the 
galaxy. The $(B-V)_{T}^{0}$ colour is redder than 
that corresponding to its morphological type. RC3 reports 
total asymptotic $B$ and $V$ magnitudes with associated errors 
of 0.1 mag that imply a $(B-V)$ colour in agreement to our 
observed value. The I/A class for the pair is NI.

{\bf KPG347A}. The components in this pair show an
apparent overlapping but similar morphological types.
The galaxy is classified as SA(rs)bc. Our sharp/filtered and 
$(B-I)$ images show a prominent bulge region and wrapped 
spiral arms that resemble an inner ring. Blue knotty 
features are appreciated along the arms. We classify this galaxy
as SA(r)bc. The $(B-V)_{T}^{0}$ colour is representative of Sa 
types. Our EE class is 12.

{\bf KPG347B}. The galaxy is classified as SA(rs)bc. Our 
sharp/filtered and $(B-I)$ images show an inclined galaxy with 
a bright nucleus and knotty features along multiple spiral 
arms. We classify this galaxy as Sc. The estimated 
$(B-V)_{T}^{0}$ colour is representative of Sa types. The 
I/A class for the pair is DI. Our EE class is 3.

{\bf KPG389A}. The components in this pair have similar 
morphological features (cf. KPG295) and their arms overlap
in the outer region. The galaxy is classified as SAB(s)b pec 
and our $(B-I)$ and sharp/filtered images show a very definite
nucleus and an adjacent small spiral arms enclosed 
by two external arms. They are knotty and resemble an inner
pseudo-ring but become diffuse at large radii. The east arm 
is seen interpenetrating to the west arm of its companion 
such that an x-like feature is formed. We classify this
galaxy as SA(rs)bpec. The $(B-V)_{T}^{0}$ colour is 
representative of Sb-Sbc types. Our EE class is 6.

{\bf KPG389B}. The galaxy is classified as SA(s)b: pec and 
our $(B-I)$ and sharp/filtered images show an elongated nuclear
region from which two long and diffuse arms emerge. The west arm 
appears crossing the east arm of its companion forming an x-shaped 
feature. We classify this galaxy as SABb pec. The $(B-V)_{T}^{0}$ 
colour is representative of Sa-Sab types. The colour profiles
show a tendency to be flat after 20$\arcsec$. The I/A class for 
the pair is LI. Our EE class is 7.

{\bf KPG396A}.  The galaxy is classified as SB(s)d: sp. Our 
$(B-I)$ and sharp/filtered images show an apparently inclined 
system with an elongated nucleus and a spiral pattern that 
is difficult to trace. The $(B-V)_{T}^{0}$ colour is more
representative of Sm-Im types.

{\bf KPG396B}.  The galaxy is classified as SB(s)d. Our $(B-I)$
and sharp/filtered images show a prominent bar structure
and a complex, knotty spiral pattern with differing pitch 
angles. The $(B-V)_{T}^{0}$ colour is representative of a Sm-Im
types. The colour profiles show structure and a tendency to be 
flat after 25$\arcsec$. The I/A class for the pair is DI.
Our EE class is 1.

{\bf KPG404A}. The pair show an apparent low degree of 
overlapping. The galaxy is classified as SB(s)b pec. 
Our sharp/filtered and $(B-I)$ images show a prominent nucleus 
and a faint adjacent linear bar from which
two spiral arms (integral sign tidal arms) emanate. 
The arms are wrapped and simulate an inner ring structure.
A third small galaxy can be appreciated near the end of the 
western arm (tail). The southern arm is seen forming a bridge 
to its companion galaxy. We classify this galaxy as SB(r)b pec.
The $(B-V)_{T}^{0}$ colour is representative of Sab-Sb types. 
The colour profiles show a tendency to be flat after 25$\arcsec$.
Our EE class is 10.

{\bf KPG404B}. The galaxy is classified as SA(s)b pec. Our 
sharp/filtered and $(B-I)$ images show a bright nucleus and  
apparently strong dust lanes along the arms (see the arm 
towards the companion galaxy). The $(B-V)_{T}^{0}$ colour is 
representative of Sab-Sb types. The I/A class for the pair is LI. 
Our EE class is 11.

{\bf KPG426A}. The galaxy is classified  as S?. Our $(B-I)$ 
and sharp/filtered images show a prominent bulge and barred
structure. At either end of the bar, strong condensations
and two thin/faint and wrapped arms emanate. There is also 
indication of a ring enclosing the bar. We classify this
galaxy as SB(r)b. The $(B-V)_{T}^{0}$ colour is representative
of S0-S0a types. The colour profiles show a tendency to be flat 
after 25$\arcsec$. Our EE class is 10.

{\bf KPG426B}. The galaxy is classified  as S?.  Our $(B-I)$ 
and sharp/filtered images show a prominent but elongated bulge 
region and a very faint spiral pattern. An unsharp masking image 
suggest that an arc-like structure may be present at the western
external part. We classify this galaxy as SABb. The $(B-V)_{T}^{0}$ 
colour is representative of S0a-Sa types. The I/A class for the pair 
is DI. Our EE class is 8.

{\bf KPG440A}. The galaxy is classified as SAB(rs)d and our 
$(B-I)$ and sharp/filtered images show a bright central 
bar and a multiple spiral pattern with knotty structure
all along the arms. We classify this galaxy as SBcd. The 
$(B-V)_{T}^{0}$ colour is definitely bluer than that 
corresponding to its morphological type. Our EE class is 6.

{\bf KPG440B}. The galaxy is classified as SBc? and our 
$(B-I)$ and sharp/filtered images show a very complex dust 
lane structure along the plane of the galaxy, similar to 
KPG332b. In our images it is difficult to trace the nuclear 
region, bulge or bar. The $(B-V)_{T}^{0}$ colour is 
representative of E-S0 types. The I/A class for the pair is 
NI.

{\bf KPG455A}. The galaxy is classified as SB(s)b and our 
$(B-I)$ and sharp/filtered images show a bright nuclear
region and two faint wrapped spiral arms resembling an inner 
ring. It is difficult to trace the presence of a bar, although
the bulge region is elongated. The spiral pattern becomes diffuse at 
the outer parts. We classify this galaxy as SA(r)b. The 
$(B-V)_{T}^{0}$ colour is representative of S0a-Sa types.

{\bf KPG455B}. The galaxy is classified as SB(s)bc but our 
$(B-I)$ and sharp/filtered images show a bright nucleus from 
which a spiral pattern emerge. Blue knotty structure is observed 
along the arms. At our resolution, we have no clear evidence of a 
barred structure. We classify this galaxy as  SA(s)bc. The 
$(B-V)_{T}^{0}$ colour is representative of Sc-Scd types. The 
I/A class for the pair is DI. Our EE class is 7.

\subsection{Results}

Table \ref{finalresult} is a summary of the results found in this
work. Column (1) gives the pair catalogued number, Column (2) gives
the Hubble Type as reported in NED, Column (3) gives the Hubble Type
as estimated in this work, Column (4) gives the
Elmegreen (EE) class, Column (5) gives the revised Karachentsev 
interaction I/A class, Column (6) remarks when a flat colour
profile is present, and finally Column (7) remarks the presence
of Bars, Knots, Rings and Shell structures.

\begin{table*}
  \caption[]{Final Results from this Study}
\label{finalresult}
\begin{tabular}{cccccccc}
\hline\noalign{\smallskip}
Galaxy pair & HUBBLE TYPE(NED) & HUBBLE TYPE(THIS WORK) & EE CLASS & I/A CLASS & PROFILE & NOTES \\
\hline\noalign{\smallskip}
KPG64A &    SA(s)b pec &        Sc pec  &              6  &         LI   &        &    K   \\   
KPG64B &    SB(s)a pec &        Sbc     &              6  &              &     F  &    K \\
KPG68A &    Scd:       &                &              3  &         DI   &        &    K \\
KPG68B &    SBb:       &                &             10  &              &     F  &    B,K \\
KPG75A &    E?         &        Sab     &                 &         DI   &        &         \\
KPG75B &    SB?        &        Sb      &              7  &              &        &        \\
KPG88A &    SA(s)c     &                &             12  &         DI   &     F  &    K  \\
KPG88B &    SBcd:      &                &             10  &              &        &    B \\
KPG98A &    Scd:       &                &              2  &         DI   &     F  &    K \\
KPG98B &    S?         &        (R)Sa   &                 &              &        &    R \\
KPG102A &   Sa         &        SBab    &                 &         LI   &        &    B \\
KPG102B &   Sb         &        Sc      &             11  &              &     F  &  \\
KPG103A &   Sc         &                &                 &         DI   &     F  &   \\
KPG103B &   Sa         &                &                 &              &        & \\
KPG108A &   Sbc        &                &                 &         NI   &     F  &  \\
KPG108B &   Sb         &                &                 &              &     F  &  \\
KPG112A &   S0/a       &        Sa      &                 &         LI   &     F  &  \\
KPG112B &   S0:        &        Sa      &             12  &              &        & \\
KPG125A &   Pec        &        Sab pec &                 &         DI   &     F  &    \\  
KPG125B &   S pec      &        Sc pec  &              9  &              &     F  &    K \\
KPG136A &   S?         &        Sbc     &             10  &         DI   &        & \\
KPG136B &   S?         &        S(r)ab  &              8  &              &        &    R \\
KPG141A &   S?         &        Sbc     &                 &         DI   &        &    K \\
KPG141B &   S?         &        Sb      &              7  &              &        & \\
KPG150A &   Sa         &        S(r)b   &              7  &         NI   &        &    R \\
KPG150B &   SBb        &        SB(r)c  &              8  &              &        &    B,K,R \\
KPG151A &   Sc         &                &                 &         LI   &        & \\
KPG151B &   SB?        &        SBb     &              8  &              &     F  &    B \\
KPG156A &   SA(r)c pec &                &             11  &         LI   &        &    K,R \\
KPG156B &   SB(rs)c pec&        SBbc pec&                 &              &     F  &    B \\
KPG159A &   Sb         &                &                 &         DI   &        & \\
KPG159B &   Sb         &        SBb     &                 &              &        &    B \\
KPG160A &   SB(s)a     &        (R')SB(s)a&            8  &         DI   &     F  &    B,R \\
KPG160B &   SBa        &        Sb      &                 &              &        &    K \\
KPG168A &   Sa         &        S(s)b   &              6  &         LI   &     F  &    R  \\
KPG168B &   Sc         &                &                 &              &     F  &    K \\
KPG195A &   SB(s)a pec &        SB(s)b pec&            6  &         DI   &     F  &    B  \\
KPG195B &   SB(s)m?    &        Sc      &                 &              &        &    K \\
KPG211A &   (R')SAB(rs)ab pec:&         &              8  &         DI   &     F  &    B  \\
KPG211B &   Sb:        &        Sa      &              7  &              &        &   \\
KPG216A &   SB(s)b pec:&                &             10  &         DI   &        &    B \\
KPG216B &   SAB(s)c pec:&       SBc pec &              9  &              &     F  &    B \\
KPG249A &   SAB(rs)cd pec&      SABcd pec&             6  &         LI   &     F  &    B,K \\ 
KPG249B &   IBm pec    &        SBm pec  &                &              &     F  &    B,K \\
KPG295A &   SAB(rs)a pec&       (R')SAB(r)a pec&       8  &         DI   &        &    B,R   \\  
KPG295B &   SAB(rs)ab pec&               &             8  &              &        &    B,R\\
KPG302A &   SAB(rs)c:  &        Sc       &            12  &         DI   &        &    K \\
KPG302B &   SB0/a pec  &        SBbc pec &             6  &              &     F  &    B \\
KPG313A &   SAB(rs)cd  &                 &             1  &         DI   &     F  &    B,R,K \\
KPG313B &   Sab:sp     &                 &                &              &        & \\
KPG332A &   SA(rs)c    &        Sc       &             3  &         NI   &        &    K \\
KPG332B &   Sc: sp     &                 &                &              &        & \\
KPG347A &   SA(rs)bc   &                 &            12  &         DI   &        &    K \\ 
KPG347B &   SA(rs)bc   &        Sc       &             3  &              &        &    K \\
\hline\noalign{\smallskip}
\end{tabular}
F = Flat Colour Profile \\
B = Bar \\
K = Presence of Knots \\
R = Ring  \\
Sh = Shell  \\
\end{table*}

\setcounter{table}{3}

\begin{table*}
\caption[]{Continued.}
\label{finalresult}
\begin{tabular}{cccccccc}
\hline\noalign{\smallskip}
Galaxy pair & HUBBLE TYPE(NED) & HUBBLE TYPE(THIS WORK) & EE CLASS & I/A CLASS & PROFILE & NOTES \\
\hline\noalign{\smallskip}
KPG389A &   SAB(s)b pec&        SA(rs)b pec&           6  &         LI   &        &    K \\
KPG389B &   SA(s)b: pec&        SABb pec &             7  &              &     F  &    B \\
KPG396A &   SB(s)d: sp &                 &                &         DI   &        &    B \\
KPG396B &   SB(s)d     &                 &             1  &              &     F  &    B,K \\
KPG404A &   SB(s)b pec &       SB(r)b pec&            10  &         LI   &     F  &    B,R \\
KPG404B &   SA(s)b pec &                 &            11  &              &        &  \\
KPG426A &   S?         &       SB(r)b    &            10  &         DI   &     F  &    B,R \\
KPG426B &   S?         &       SABb      &             8  &              &        &    B,Sh \\
KPG440A &   SAB(rs)d   &       SBcd      &             6  &         NI   &        &    B,K  \\
KPG440B &   SBc?       &                 &                &              &        &    B  \\
KPG455A &   SB(s)b     &       SA(r)b    &                &         DI   &        &    R  \\
KPG455B &   SB(s)bc    &       SA(s)bc   &             7  &              &        &    K  \\
\hline\noalign{\smallskip}
\end{tabular}
F = Flat Colour Profile \\
B = Bar \\
K = Presence of Knots \\
R = Ring  \\
Sh = Shell  \\
\end{table*}

A reclassification of the galaxies in the (S+S) sample
was made using our CCD data. The original classifications 
were made on the low resolution POSS.  We revise the Hubble
classifications for at least one component in  44\% of our 
pairs (29 galaxies). An appreciable change in Hubble types 
$\Delta T \geq 2$ was found in 25 galaxies. The bulk of the
sample is comprised of (S+S) pairs. Very few galaxies could 
be classified as irregular, and they may well be severely 
distorted spirals. If we consider $\Delta T \geq 2$ as a 
minimum value for morphological discordance in pairs, then 
half of our pairs show morphological concordance between 
pair members; 17 pairs (51 \%) show $\Delta T \sim 1$ and 
16 pairs (49 \%) $\Delta T \geq 2$. This could explain, in 
part, the strong correlation found between $(B-V)$ colour 
indices (Holmberg Effect) between members of this sample.

Elmegreen \& Elmegreen (\cite{elmegreenelmegreen82})
developed a 12-division morphological system to classify 
spiral galaxies according to the regularity of their spiral 
arm structure. This spiral arm classification correlates with
the presence of density waves as in grand design galaxies.
Following that work, we succeeded in classifying 43 spirals.
Some of the spirals in the global sample are nearly edge-on,
strongly interacting or simply do not fit into the
Elmegreen \& Elmegreen classes. From 26 barred spirals, 18 
are grand design and only 2 are flocculent. From 40 non-barred 
galaxies, 17 are grand design and 5 are flocculent. Grand 
design structure seems to be connected with binary galaxies, 
but strongly for barred than for non-barred galaxies. These
results seem to be consistent with those of Elmegreen \& 
Elmegreen (\cite{elmegreenelmegreen82}). We also have found 
knotty features in 24 galaxies and have detected rings or 
pseudo-ring features in 13 galaxies.

Interestingly, a fraction of the spirals have ``open
arms'' that could be interpreted in the framework of the 
simulations of Noguchi (\cite{noguchi90} and references therein). 
The simulated galaxies, have similar sizes as those in our sample.
Noguchi's models follow both the stellar and gaseous components
evolution in a disc galaxy during the encounter. Briefly
speaking, in this scenario of moderate interactions,
the bar develops quite soon and it is long lasting, while the
ring develops later and the gas follows the configuration of the
ring. Four different phases may be evidenced: 1) open arms
appear (integral sign) after perigalacticon, 2) a bar develops,
the arms start to close and the gas start to follow the star
configuration, 3) the arms are completely closed around the bar
and form a ring; the gas is mainly concentrated in the center
and ring, and 4) the ring starts to be disrupted by the dynamics
and the overall appearance of the galaxy becomes nearly asymmetric.

As noted above, a fraction of the spirals in our sample
(40\%/20\%) show (bar/ring) features which could be a 
transient phenomenon of the interaction in Noguchi's models. 
The bars are always redder while the rings and knots are always 
bluer compared with the galaxy outskirts. Most of the bridges 
and tails maintain the colour of the outskirts of the galaxies. 
The knotty structure along the arms and disks, confirms the 
global nature of the star formation induced by the interactions.  
From 33 pairs, 20 can be classified according to the I/A
class DI, 9 pairs as LI, and 4 pairs as NI. We have not 
detected any AT class, perhaps as a selection bias from our
observing strategy. The sequence AT-LI-DI-NI has been 
interpreted as a sequence going from strongest to weak for 
tidal distortion or from most to least dynamically evolved.   
According to this, our (S+S) pairs are mainly involved in 
interactions of moderate level.

An interesting correlation has been found between the optical
morphology and the global photometric properties in these
pairs. From 26 barred galaxies, 15 show a flat behaviour 
(negligible gradient) in the azimuthally averaged colour 
profile while from 40 non-barred galaxies, 12 show flat colour 
profiles. This result may indicate that the bar acts unifying 
the stellar populations of the bulge and disk, in agreement 
with a secular evolutionary scenario, and consistent 
with the results in Gadotti \& dos Anjos 
(\cite{gadottidosanjos01}) and Zaritsky et al. 
(\cite{zaritskyetal94}) where barred spiral galaxies have 
flatter abundance gradients than unbarred spirals.

\section{Conclusions}
\label{conclusion}

In order to analyze the photometrical signature of gravitational 
interactions in spiral galaxies, we present results of our 
$BVRI$ surface photometry for a first set of 33 (S+S) pairs 
from the Karachentsev (\cite{kara72}) catalogue. We show that
our derived parameters are generally in good agreement with
those reported in RC3, aperture photometry catalogues and 
other individual photometric works. In addition, we present
multiaperture photometry in order to facilitate further 
comparisons and contribute to the existing database of aperture 
photometry. The combination of 2D $(B-I)$ colour maps and 
sharp/filtered $B$ images appears to be a powerful technique 
both for morphological classification and for revealing fine 
structural details most likely related to encounters that are in
various early and late stages. There is a tendency of barred 
galaxies to show grand design morphologies and flat colour 
profiles. In general our data suggest that our sample is 
undergoing moderate interactions which appear to be adequate
to stimulate a nonaxisymmetric potential that generate a global 
response as evidenced by the presence of bars, rings, pseudo-rings 
and knotty structures along the arms and disks of the spiral
galaxies in (S+S) pairs.

\begin{acknowledgements}

I.P. thanks the staff of the Observatorio Guillermo Haro for the 
help in the observations and also likes to thank E. Athanassoula for 
interesting discussions on spiral structure. These discussions were 
made possible via the ECOS/ANNUIES exchange project M99-U02, for 
which we are thankful. H.H.T. thanks J. Sulentic for his valuable
comments. H.H.T. and I.P. thank the referee  Dr. R. Rampazzo for his 
careful reading of this manuscript. His constructive criticism has 
helped us to clarify and focus our paper enormously.

\end{acknowledgements}

\newpage

\appendix

\section{Aperture Photometry}

Since the birth of galaxy photometry (Whitford 
\cite{whitford36}), the amount of photometric data has 
increased exponentially (Prugniel \cite{prugniel87}). 
However, this data is inhomogeneous both in quality and 
format: photographic, photoelectric or more recently, CCD 
observations. The data are usually presented as centered 
aperture photometry through circular or elliptical apertures 
or as photometric profiles. In order to take into account 
the continuously growing amount of photometric data and 
at the same time, to make different photometric data 
reports somehow comparable, we present in Table 
\ref{tabmagapert} our estimations of integrated magnitudes 
in three concentric circular apertures. Columns (2) and (3) 
give the logarithm of the aperture radius (in arcmin) for 
pair component (A) and (B). Columns (4)-(11) give their 
corresponding magnitudes in $B$, $V$, $R$ and $I$ bands, 
respectively. The small difference in aperture sizes 
suggest that the contribution of the sky to the errors 
in the magnitudes is relatively small. Typical 
uncertainties in the magnitudes are 0.15, 0.14, 
0.15 and 0.14 in $B$, $V$, $R$ and $I$ bands 
respectively.

\begin{table*}
\caption[]{Magnitudes at Different Circular Apertures}
\label{tabmagapert}
\begin{tabular}{ccccccccccc}

\hline\noalign{\smallskip}
KPG & \multicolumn{2}{c}{Log(Aperture)} & \multicolumn{2}{c}{$B$}  &
\multicolumn{2}{c}{$V$} & \multicolumn{2}{c}{$R$} & \multicolumn{2}{c}{$I$} \\
\cline{2-11}
Number & Cpt A & Cpt B & Cpt A & Cpt B & Cpt A & Cpt B & Cpt A & Cpt B &
Cpt A & Cpt B \\
\hline\noalign{\smallskip}
KPG64 &1.602&1.301&14.011&15.608&13.215&14.681&12.545&13.976&11.938&13.374\\
     &1.698&1.477&13.789&15.230&13.043&14.335&12.378&13.645&11.780&13.079\\
     &1.778&1.602&13.647&14.990&12.934&14.104&12.268&13.408&11.685&12.863\\
KPG68 &1.301&1.301&14.490&15.288&13.279&14.089&12.738&13.578&11.983&12.859 \\
      &1.477&1.477&14.047&14.944&12.881&13.726&12.364&13.231&11.640&12.527 \\
      &1.602&1.602&13.833&14.669&12.673&13.433&12.172&12.949&11.463&12.258 \\
KPG75 &1.301&1.301&14.905&15.695&13.933&14.765&13.394&14.202&12.320&13.109\\
     &1.477&1.477&14.761&15.504&13.801&14.583&13.265&14.024&12.202&12.954\\
     &1.602&1.602&14.691&15.387&13.728&14.465&13.192&13.908&12.145&12.864\\
KPG88 &1.176&1.176&16.318&16.463&15.063&15.331&14.263&14.605&13.394&13.806\\
     &1.397&1.397&15.648&16.069&14.414&14.912&13.633&14.201&12.786&13.393\\
     &1.544&1.544&15.260&16.028&14.034&14.787&13.262&14.078&12.438&13.256\\
KPG98 &1.397&1.397&15.580&15.769&13.859&14.507&13.094&13.802&12.163&12.962\\
     &1.544&1.544&15.552&15.641&13.771&14.568&12.983&13.789&12.070&12.985\\
     &1.653&1.653&15.589&15.602&13.719&14.917&12.909&13.964&12.018&13.206\\
KPG102 &1.000&1.000&16.255&16.084&15.560&15.320&15.070&14.817&14.202&13.914\\
     &1.301&1.301&15.903&15.425&15.208&14.773&14.753&14.337&13.882&13.471\\
     &1.477&1.477&15.703&15.266&14.992&14.575&14.583&14.170&13.731&13.314\\
KPG103 &1.176&1.176&16.378&15.411&15.557&14.665&15.051&14.246&14.218&13.469\\
     &1.301&1.301&16.256&15.311&15.485&14.582&14.987&14.179&14.157&13.394\\
     &1.397&1.397&16.189&15.239&15.477&14.521&14.976&14.136&14.154&13.343\\
KPG108 &1.653&1.653&15.002&14.344&13.979&13.293&13.477&12.765&12.442&11.719\\
     &1.740&1.740&14.962&14.131&13.933&13.064&13.439&12.554&12.395&11.517\\
     &1.812&1.812&14.957&14.077&13.889&12.982&13.403&12.478&12.354&11.444\\
KPG112 &1.000&1.000&13.891&14.101&12.973&13.161&12.391&12.550&11.678&11.817 \\
       &1.176&1.176&13.684&13.868&12.772&12.942&12.188&12.333&11.481&11.596 \\
       &1.301&1.301&13.561&13.716&12.657&12.798&12.068&12.192&11.365&11.454 \\
KPG125 &1.477&1.477&13.967&13.248&13.038&12.382&12.314&11.751&11.506&10.956\\
     &1.602&1.602&13.871&13.082&12.949&12.198&12.218&11.574&11.417&10.773\\
     &1.698&1.698&13.806&13.041&12.882&12.134&12.146&11.514&11.357&10.709\\
KPG136 &1.301&1.301&14.751&15.019&14.263&14.435&13.694&13.812&13.211&13.282\\
     &1.397&1.397&14.688&14.924&14.213&14.341&13.649&13.728&13.182&13.206\\
     &1.477&1.477&14.657&14.871&14.192&14.286&13.632&13.677&13.183&13.168\\
KPG141 &1.602&1.602&14.767&15.205&13.934&14.206&13.460&13.784&12.612&12.883\\
     &1.698&1.698&14.670&14.905&13.882&13.924&13.402&13.525&12.618&12.597\\
     &1.778&1.778&14.629&14.818&13.885&13.841&13.391&13.464&12.687&12.497\\
KPG150 &1.477&1.477&14.458&13.931&13.401&12.815&12.614&12.110&11.901&11.311\\
     &1.602&1.602&14.282&13.626&13.267&12.532&12.486&11.860&11.791&11.061\\
     &1.698&1.698&14.207&13.423&13.234&12.344&12.443&11.701&11.777&10.888\\
KPG151 &1.301&1.301&15.966&15.410&14.836&14.308&14.063&13.587&13.296&12.883\\
     &1.477&1.477&15.684&15.090&14.543&13.988&13.774&13.268&13.010&12.567\\
     &1.602&1.602&15.419&14.929&14.298&13.808&13.541&13.083&12.781&12.388\\
KPG156 &1.301&1.301&14.320&14.928&13.630&14.291&13.064&13.752&12.521&13.271 \\
       &1.477&1.477&13.820&14.783&13.184&14.094&12.639&13.558&12.130&13.062 \\
       &1.653&1.653&13.491&14.701&12.876&13.887&12.347&13.369&11.851&12.842 \\
KPG159 &1.000&1.301&16.881&15.553&16.230&14.671&15.711&14.056&15.231&13.445 \\
       &1.301&1.477&16.657&14.938&15.993&14.134&15.456&13.554&14.978&12.978 \\
       &1.477&1.602&16.515&14.755&15.839&13.979&15.276&13.410&14.771&12.843 \\
KPG160 &1.544&1.544&13.966&15.295&13.142&14.448&12.641&14.004&11.750&13.012\\
     &1.602&1.602&13.932&15.269&13.098&14.391&12.598&13.955&11.706&12.939\\
     &1.653&1.653&13.911&15.240&13.067&14.332&12.565&13.907&11.673&12.861\\
KPG168 &1.602&1.301&13.040&15.461&12.061&14.701&11.436&14.178&10.811&13.678\\
     &1.740&1.477&12.928&15.249&11.951&14.412&11.327&13.907&10.716&13.373\\
     &1.845&1.602&12.855&15.136&11.876&14.206&11.253&13.713&10.659&13.151\\
KPG195 &1.698&1.698&13.241&14.352&12.472&13.735&11.869&13.245&11.225&12.697\\
     &1.778&1.778&13.175&14.324&12.408&13.678&11.810&13.185&11.173&12.638\\
     &1.845&1.845&13.120&14.291&12.358&13.607&11.762&13.107&11.136&12.564\\
\hline\noalign{\smallskip}
\end{tabular}
\end{table*}

\setcounter{table}{0}
\begin{table*}
  \caption[]{Continued.}
\begin{tabular}{ccccccccccc}
\hline\noalign{\smallskip}
KPG & \multicolumn{2}{c}{Log(Aperture)} & \multicolumn{2}{c}{$B$}  &
\multicolumn{2}{c}{$V$} & \multicolumn{2}{c}{$R$} & \multicolumn{2}{c}{$I$}\\
\cline{2-11}
Number & Cpt a & Cpt b & Cpt a & Cpt b & Cpt a & Cpt b & Cpt a & Cpt b &
Cpt a & Cpt b \\
\hline\noalign{\smallskip}
KPG211 &1.477&1.176&13.990&15.928&12.954&14.801&12.274&14.111&11.604&13.406\\
     &1.602&1.397&13.773&15.729&12.752&14.604&12.077&13.918&11.416&13.200\\
     &1.698&1.544&13.696&15.690&12.674&14.538&11.996&13.851&11.341&13.112\\
KPG216 &1.000&1.698&14.785&13.373&14.556&12.955&14.163&12.622&13.724&12.067\\
     &1.301&1.778&14.111&13.327&13.917&12.848&13.550&12.527&13.127&11.962\\
     &1.477&1.845&13.720&13.313&13.587&12.766&13.262&12.462&12.886&11.881\\
KPG249 &1.301&1.301&13.115&13.469&12.763&13.090&12.303&12.597&11.963&12.267 \\
       &1.477&1.477&12.715&13.215&12.388&12.814&11.942&12.326&11.615&11.973 \\
       &1.602&1.602&12.554&13.046&12.221&12.625&11.774&12.142&11.455&11.776 \\
KPG295 &1.477&1.477&13.754&13.708&12.932&12.875&12.288&12.251&11.730&11.634\\
     &1.602&1.602&13.569&13.500&12.747&12.679&12.113&12.071&11.564&11.466\\
     &1.698&1.698&13.418&13.338&12.595&12.528&11.963&11.934&11.431&11.358\\
KPG302 &1.929&1.477&11.255&14.076&10.740&13.675&10.332&13.346& 9.619&12.659\\
     &2.000&1.698&11.207&13.862&10.688&13.422&10.281&13.070& 9.575&12.362\\
     &2.060&1.845&11.178&13.827&10.651&13.337&10.243&12.957& 9.547&12.235\\
KPG313 &1.653&1.653&13.409&13.351&12.765&12.224&12.249&11.443&11.821&10.676\\
     &1.778&1.778&13.184&13.259&12.554&12.141&12.049&11.364&11.655&10.600\\
     &1.875&1.875&13.107&13.223&12.482&12.105&11.990&11.325&11.648&10.558\\
KPG332 &1.698&1.698&12.512&13.556&11.703&12.434&11.297&11.889&10.375&10.740\\
     &1.845&2.000&12.252&12.737&11.455&11.657&11.056&11.151&10.151&10.035\\
     &1.954&2.176&12.169&11.972&11.352&10.988&10.952&10.523&10.063& 9.446\\
KPG347 &1.477&1.477&13.008&13.033&12.259&12.087&11.726&11.422&11.084&10.648 \\
       &1.602&1.602&12.616&12.669&11.859&11.747&11.327&11.102&10.690&10.349 \\
       &1.698&1.698&12.332&12.405&11.564&11.492&11.027&10.864&10.387&10.125 \\
KPG389 &1.477&1.477&13.773&13.920&13.142&13.158&12.528&12.527&12.104&11.962\\
     &1.602&1.602&13.693&13.782&13.053&13.012&12.436&12.386&12.024&11.810\\
     &1.698&1.698&13.659&13.744&12.997&12.930&12.373&12.302&11.977&11.709\\
KPG396 &1.602&1.602&14.939&14.138&14.344&13.719&13.946&13.403&13.133&12.581\\
     &1.698&1.698&14.852&14.087&14.221&13.629&13.822&13.310&12.979&12.421\\
     &1.778&1.778&14.754&14.063&14.098&13.560&13.700&13.238&12.825&12.278\\
KPG404 &1.477&1.698&14.070&12.796&13.393&12.043&12.846&11.474&12.196&10.785 \\
       &1.602&1.812&13.935&12.624&13.254&11.875&12.714&11.314&12.086&10.635 \\
       &1.698&1.903&13.661&12.533&12.984&11.787&12.462&11.231&11.862&10.564 \\
KPG426 &1.301&1.301&15.059&14.759&14.053&14.888&13.406&13.918&12.792&13.299\\
     &1.397&1.397&14.941&15.066&13.942&14.180&13.298&13.568&12.689&13.024\\
     &1.477&1.477&14.871&15.018&13.870&14.122&13.222&13.511&12.620&12.969\\
KPG440 &1.698&1.929&13.535&12.523&13.045&11.445&12.699&10.940&12.054& 9.868\\
     &1.845&2.000&13.190&12.418&12.823&11.335&12.481&10.842&11.996& 9.765\\
     &1.929&2.079&13.018&12.348&12.764&11.248&12.416&10.780&12.080& 9.692\\
KPG455 &1.397&1.740&14.124&13.348&13.302&12.617&12.792&12.125&11.898&11.203\\
     &1.544&1.812&14.018&13.259&13.185&12.540&12.676&12.054&11.793&11.127\\
     &1.653&1.875&14.070&13.195&13.140&12.482&12.613&11.999&11.742&11.063\\
\hline\noalign{\smallskip}
\end{tabular}
\end{table*}


\begin{thebibliography}{}

\bibitem[2000]{avilafirmani00}
     Avila-Reese, V., \& Firmani, C. 2000, 
     Rev. Mex. Astron. Astrof\'\i s., 36, 23

\bibitem[1998]{avilaetal98}
     Avila-Reese, V., Firmani, C., \& Hern\'andez, X. 1998,
     ApJ, 505, 37

\bibitem[1996]{baughetal96}
     Baugh, C.M., Cole, S., \& Frenk, C.S. 1996, 
     MNRAS, 283, 1361 

\bibitem[1995]{bergvalljohansson95}
    Bergvall, N., \& Johansson, L. 1995,
    A\&AS, 113, 499


\bibitem[1982]{bursteinheiles82}
    Burstein, N., \& Heiles, K. 1982,
    AJ, 87, 1165

\bibitem[1989]{cardellietal89}
    Cardelli, J.A., Clayton, G.C., \& Mathis, J.S. 1989,
    ApJ, 345, 245

\bibitem[1991]{chevalierilovaisky91}
    Chevalier, C., \& Ilovaisky, S.A.  1991,
    A\&AS, 90, 225

\bibitem[1994]{dejongvanderkruit94}
    de Jong, R., \& van der Kruit, P.C. 1994,
    A\&AS, 106, 451

\bibitem[1991]{devaucouleursetal91}
    de Vaucouleurs, G., de Vaucouleurs, A., Corwin, H.G., et al. 1991,
    Third Reference Catalogue of Bright Galaxies, New York,
    Springer-Verlag

\bibitem[1988]{devaucouleurslongo88}
    de Vaucouleurs, A., \& Longo, G. 1988,
    Catalogue of visual and infrared photometry of galaxies from 0.5
    micrometer to 10 micrometer, University of Texas

\bibitem[1979]{doroshenkoterebizh79}
    Doroshenko, V.T., \& Terebizh, V.Y. 1979,
    Soviet Astron. Lett., 5, 305


\bibitem[1997]{dultzin97}
    Dultzin-Hacyan, D. 1997, 
    Rev. Mex. Astron. Astrof\'\i s., 6, 132


\bibitem[1982]{elmegreenelmegreen82}
    Elmegreen, B.G., \& Elmegreen, D.M. 1982,
    MNRAS, 201, 1021

\bibitem[2001]{gadottidosanjos01}
    Gadotti, D.A., \& dos Anjos, S. 2001,
    AJ, accepted (astro-ph/0106303)


\bibitem[1997]{giovanellietal97}
    Giovanelli, R., Haynes, M.P., da Costa L.N., Freudling, W.,
    Salser, J.J., \& Wegner, G. 1997,
    ApJ, 477, 1

\bibitem[1977]{godwinetal77}
    Godwin, J.G., Bucknell, M.J., Dixon, K.L., Green, M.R., Peach, J.V.,
    \& Wallis, R.E. 1977,
    The Observatory, 97, 238

\bibitem[1992]{han92}
    Han, M. 1992,
    ApJS, 81, 35

\bibitem[1984]{haynesgiovanelli84}
    Haynes, M.P., \& Giovanelli, R. 1984,
    AJ, 89, 758

\bibitem[1999]{toledoetal99}
    Hern\'andez-Toledo, H.M., Dultzin-Hacyan D., Gonz\'alez, J.J.,
    \& Sulentic, J. 1999,
    AJ, 118, 108

\bibitem[1958]{holmberg58}
    Holmberg, E. 1958,
    Lund Medd. Astron. Obs. Ser. II, 136, 1


\bibitem[2000]{jansenetal00}
     Jansen, R.A., Franx, M., Fabricant, D., \& Caldwell, N. 2000,
     ApJS, 126, 271


\bibitem[1990]{johanssonbergvall90}
    Johansson, L., \& Bergvall, N. 1990,
    A\&AS, 86, 167


\bibitem[1998]{junquieraetal98}
    Junqueira, S., de Mello, D.F., \& Infante, L. 1998,
    A\&AS, 129, 69

\bibitem[1972]{kara72}
    Karachentsev, I.D. 1972,
    Catalogue of Isolated Pairs of Galaxies in the Northern
    Hemisphere, Comm. Spec. Ap. Obs., 7, 1


\bibitem[2001]{kauffmannetal01}
    Kauffmann, G., Charlot, S., \& Balogh, M.L. 2001, 
    MNRAS, submitted (astro-ph/0103130)


\bibitem[1993]{kauffmannetal93}
    Kauffmann, G., White, S.D.M., \& Guiderdoni, B. 1993, 
    MNRAS, 264, 201

\bibitem[1988]{keel88}
    Keel, W. C. 1988,
    A\&AS, 202, 41

\bibitem[1987]{kennicuttetal87}
    Kennicutt, R.C., Roettiger, K.A., Keel, W.C., van der Hulst, J.M.,
    \& Hummel, E. 1987,
    AJ, 93, 1011

\bibitem[1991]{knapenkruit91}
    Knapen, J.H. \& van der Kruit, P.C. 1991,
    A\&A, 248, 57


\bibitem[1991]{laceysilk91}
    Lacey, C.G., \& Silk, J. 1991, 
    ApJ, 381, 14


\bibitem[1978]{larsontinsley78}
    Larson, R.B., \& Tinsley, B.M. 1978,
    ApJ, 219, 46
 

\bibitem[1998]{laurikainenetal98}
    Laurikainen, E., Salo, H., \& Aparicio, A. 1998,
    A\&AS, 129, 517

\bibitem[1996]{marquezmoles96}
    M\'arquez, I., \& Moles, M. 1996,
    A\&AS, 120, 1

\bibitem[1999]{marquezmoles99}
    M\'arquez, I., \& Moles, M. 1999,
    A\&A, 344, 421 


\bibitem[2001]{mayeretal01}
    Mayer, L., Governato, F., Colpi, M., Moore, B., Quinn, T.,
    Wadsley, J., Stadel, J., \& Lake, G. 2001, 
    ApJ, 547, L123


\bibitem[1998]{metcalfeetal98}
    Metcalfe, N., Ratcliffe, A., Shanks, T., \& Fong, R. 1998,
    MNRAS, 294, 147 


\bibitem[1996]{mooreetal96}
    Moore, B., Katz, N., Lake, G., Dressler, A., \& Oemler, A., Jr. 1996, 
    Nature, 379, 613

\bibitem[1990]{noguchi90}
    Noguchi, M. 1990,
    in Paired and Interacting Galaxies: IAU Colloquium, 124, 711


\bibitem[1987]{prugniel87}
    Prugniel Ph. 1987, 
    CDS Bull., 33, 17

\bibitem[1998]{prugnielheraudeau98}
    Prugniel, Ph., \& H\'eraudeau, Ph. 1998,
    A\&AS, 128, 299

\bibitem[1996]{reduzzirampazzo96}
    Reduzzi, L., \& Rampazzo, R. 1996,
    A\&AS, 116, 515

\bibitem[1993]{reshetnikov93}
    Reshetnikov, V.P. 1993, 
    A\&AS, 99, 257

\bibitem[1994]{robertshaynes94}
    Roberts, M.S., \& Haynes, M.P. 1994,
    ARA\&A, 32, 115

\bibitem[1993]{sofue93}
    Sofue, Y. 1993,  
    PASP, 105, 308

\bibitem[1978]{stocke78}
    Stocke, J.T. 1978,
    AJ, 83, 348

\bibitem[1976]{sulentic76}
    Sulentic, J.W. 1976,
    ApJS, 32, 171

\bibitem[1982]{vanmoorsel82}
    van Moorsel, G.A. 1982,
    PhD. Thesis, University of Groningen

\bibitem[1936]{whitford36}
    Whitford, A.E. 1936, 
    ApJ, 83, 424

\bibitem[1991]{xusulentic91}
    Xu, C., \& Sulentic, J.W. 1991,
    ApJ, 374, 407

\bibitem[1994]{zaritskyetal94}
    Zaritsky, D., Kennicutt, R.C., \& Huchra, J.P. 1994,
    ApJ, 420, 87
 

\end{thebibliography}
\end{document}